\documentclass[11pt]{article}

\usepackage{authblk}

\renewcommand\Affilfont{\small}
\usepackage[utf8]{inputenc}
\usepackage[T1]{fontenc}

\usepackage{amsmath,amssymb,amsfonts,bm}
\usepackage{mathtools}

\usepackage{graphicx}
\usepackage{tikz}
\usetikzlibrary{arrows.meta,calc}
\usepackage{placeins}
\usepackage{adjustbox}

\usepackage{booktabs}

\usepackage{geometry}
\usepackage{lineno}
\usepackage[numbers,sort&compress]{natbib}
\usepackage{bibunits}
\defaultbibliographystyle{unsrtnat}
\defaultbibliography{bibliography}

\usepackage{caption}
\usepackage{hyperref}
\hypersetup{
    colorlinks=true,
    citecolor=blue,
    linkcolor=blue,
    urlcolor=blue
}
\usepackage{doi}

\newcommand{\panelgraphic}[3][]{%
    \begin{tikzpicture}
        \node[
            inner sep=0,
            anchor=south west
        ] (img) at (0,0) {%
            \includegraphics[#1]{#2}%
        };
        \node[
            anchor=south west,
            font=\bfseries\large,
            xshift=0mm,
            yshift=1mm
        ] at (img.north west) {#3};
    \end{tikzpicture}%
}

\makeatletter
\newcommand{\resetauthblk}{%
    \setcounter{authors}{0}%
    \setcounter{affil}{0}%
    \newaffiltrue
    \gdef\AB@authlist{}%
    \gdef\AB@affillist{}%
    \gdef\AB@authors{}%
    \gdef\AB@au@str{}%
    \gdef\AB@las{}%
    \gdef\AB@lasx{\protect\Authand}%
    \gdef\AB@as{}%
    \gdef\AB@affilsep{\protect\Affilfont}%
    \gdef\@thanks{}%
}
\makeatother

\begin{document}

\begin{bibunit}

\title{\textbf{Decoupling ultrastrong Rabi splitting from quantum vacuum dressing in a chiral cavity}}

\author[1]{Ayoub El-Amrani\thanks{Correspondence: ayoub.el-amrani@um6p.ma}}
\author[1]{Nasir Boubekraoui}
\author[2,3]{Zakaria Mzaouali}
\author[1]{Houssam Sabri}
\author[4]{Herschel Rabitz}
\author[1]{Abdelouahed El Fatimy}
\author[1]{Dukhyung Lee\thanks{Correspondence: dukhyung.lee@um6p.ma}}

\affil[1]{College of Physical Sciences and Engineering,
University Mohammed VI Polytechnic,
Ben Guerir 43150, Morocco}

\affil[2]{Institut f\"ur Theoretische Physik,
Universit\"at T\"ubingen,
Auf der Morgenstelle 14,
72076 T\"ubingen, Germany}

\affil[3]{J\"ulich Supercomputing Centre,
Forschungszentrum J\"ulich, J\"ulich 52428, Germany}

\affil[4]{Department of Chemistry, Princeton University, Princeton, New Jersey 08544, USA}

\date{}

\maketitle


\begin{abstract}
Polariton formation and quantum vacuum dressing are usually regarded as intertwined
consequences of ultrastrong light--matter coupling. Here we show that they
can be selectively controlled through the relative direction of rotation of the cavity field and the matter excitation. We consider cyclotron resonance coupled to a chiral cavity supporting a single circularly polarized mode. We quantize the electromagnetic field within the cavity, derive from first principles the Hamiltonian, and investigate how mode hybridization and vacuum dressing are rendered as we tune the magnetic field.
When the cavity field co-rotates with the cyclotron motion, energy
is exchanged resonantly between light and matter, producing a large Rabi
splitting and conventional polaritons, while the ground state remains the
bare vacuum. Reversing the magnetic field makes the cyclotron motion counter-rotating with the cavity field. The same cavity mode then couples through anomalous
processes that create and annihilate light and matter excitations in pairs:
the conventional polaritonic anticrossing disappears, while the ground state
acquires squeezing and virtual excitations. Our results show that
spectral hybridization and quantum vacuum dressing are distinct manifestations
of ultrastrong coupling, offering an additional degree of freedom for engineering cavity quantum materials.
\end{abstract}



Strong light--matter coupling hybridizes electromagnetic and matter
excitations into polaritons, conventionally identified by an avoided crossing
and the associated vacuum Rabi splitting~\cite{JaynesCummings1963,Thompson1992,Weisbuch1992}. When the coupling becomes a sizeable
fraction of the bare excitation frequencies, the system enters the
ultrastrong-coupling (USC) regime, where the rotating wave approximation (RWA)
breaks down and counter-rotating and diamagnetic interactions can no longer be
neglected~\cite{CiutiPRB2005,Gunter2009,KockumNatRevPhys2019}. These terms profoundly modify the interacting ground state, giving
rise to virtual excitations, squeezing and energy shifts induced by vacuum
fields~\cite{CiutiPRB2005,DeLiberato2017Virtual}. Ultrastrong coupling
has been achieved across semiconductor and circuit-QED platforms
~\cite{Todorov2010,Niemczyk2010,FornDiaz2010}, establishing these effects as
a general feature of light--matter physics rather than a property of a
particular material realization. USC has consequently evolved from a regime
of extreme spectral hybridization into a route for manipulating matter
through vacuum electromagnetic fluctuations, with implications ranging from
transport to correlated and topological phases
~\cite{GarciaVidalCiutiEbbesen2021,Bloch2022}. Yet, in most realizations,
polariton formation and quantum vacuum dressing emerge simultaneously, making
it difficult to determine whether they are fundamentally inseparable aspects
of the same interaction~\cite{CiutiPRB2005,Gunter2009}.

Landau polaritons provide a particularly clean solid-state platform in which
to address this question. They arise from the coupling of the cyclotron
resonance (CR), a bosonic collective inter-Landau-level excitation in a
two-dimensional electron gas (2DEG), to a terahertz cavity mode. The large
collective oscillator strength of the CR and its continuous tunability with
magnetic field enabled some of the earliest demonstrations of USC in the
terahertz domain and subsequently provided access to coupling strengths approaching
and exceeding the bare transition frequency
~\cite{ScalariScience2012,Bayer2017}. Unlike systems in which the material resonance is fixed, the magnetic field tunes the matter excitation while leaving the bare photonic structure unchanged, providing a direct means of probing how light–matter coupling evolves across resonance.

Symmetry provides an additional means of reorganizing these vacuum
interactions. Symmetry breaking in matter can be encoded in the surrounding
electromagnetic fluctuations, as described by the concept of a
quantum atmosphere~\cite{Jiang2019}, while imposing chirality directly on the
cavity vacuum has been predicted to produce enantioselective energy shifts,
reshape electronic bands and generate topological phases~\cite{Hubener2021,Ke2023,Jiang2024,Yang2025}. The CR offers a particularly
direct realization of this principle because its in-plane motion has a
definite chirality that reverses with the magnetic field. Consequently, the
two circular components of the cavity field play fundamentally different
roles: the component co-rotating with the cyclotron motion mediates resonant,
particle-conserving excitation exchange, whereas the opposite component
enters through counter-rotating processes. Landau-polariton platforms thus provide a fertile setting for exploring the consequences of cavity coupling beyond the conventional polaritonic avoided crossings. The vacuum Bloch--Siegert shift
provided an experimental manifestation of this chiral asymmetry~\cite{Li2018}, and our recent work further showed that the dominant anomalous correlations,
squeezing and light--matter entanglement of multimode Landau polaritons
are predominantly associated with the counter-rotating cavity mode~\cite{El-Amrani2026}. These findings suggest that polariton formation and quantum vacuum dressing are distinct, potentially separable aspects of light–matter coupling. However, the coexistence of both circularly polarized modes in a conventional achiral cavity makes it difficult to disentangle their contributions.

Here we push this separation to the limit with a single-polarization chiral cavity. We derive the light--matter Hamiltonian for a single circularly polarized cavity mode using minimal coupling. Our analysis based on this Hamiltonian demonstrates that polariton formation and quantum vacuum dressing can be accessed independently. For one direction of the magnetic field, this mode rotates with the CR and resonantly
exchanges excitations with it: a large Rabi splitting and polariton formation
occur while the ground state remains free of virtual excitations. Reversing
the magnetic field reverses the cyclotron motion without altering the cavity
mode. The same mode then couples through processes that create and annihilate
light and matter excitations in pairs; the conventional polaritonic
anticrossing disappears while the ground state becomes squeezed and populated
by virtual excitations.

\section*{Chiral cavity with fixed circular polarization}

\begin{figure}[!t]
\centering

\newcommand{\panelAImageX}{0.0cm}
\newcommand{\panelAImageY}{-4.0cm}
\newcommand{\panelALabelX}{0.0cm}
\newcommand{\panelALabelY}{0.0cm}
\newcommand{\panelAWidth}{0.95\linewidth}

\newcommand{\panelBImageX}{0.0cm}
\newcommand{\panelBImageY}{0.0cm}
\newcommand{\panelBLabelX}{0.0cm}
\newcommand{\panelBLabelY}{0.0cm}
\newcommand{\panelBWidth}{1.00\linewidth}

\newcommand{\panelCImageX}{0.0cm}
\newcommand{\panelCImageY}{0.0cm}
\newcommand{\panelCLabelX}{0.0cm}
\newcommand{\panelCLabelY}{0.0cm}
\newcommand{\panelCWidth}{1.00\linewidth}

\newcommand{\panelDImageX}{0.0cm}
\newcommand{\panelDImageY}{0.0cm}
\newcommand{\panelDLabelX}{0.0cm}
\newcommand{\panelDLabelY}{0.0cm}
\newcommand{\panelDWidth}{1.00\linewidth}


\noindent
\begin{minipage}[t]{0.485\textwidth}
\vspace{0pt}

\makebox[0pt][l]{%
    \hspace{\panelALabelX}%
    \raisebox{\panelALabelY}[0pt][0pt]{\textbf{a}}%
}

\hspace{\panelAImageX}%
\raisebox{\panelAImageY}{%
    \includegraphics[
        width=\panelAWidth
    ]{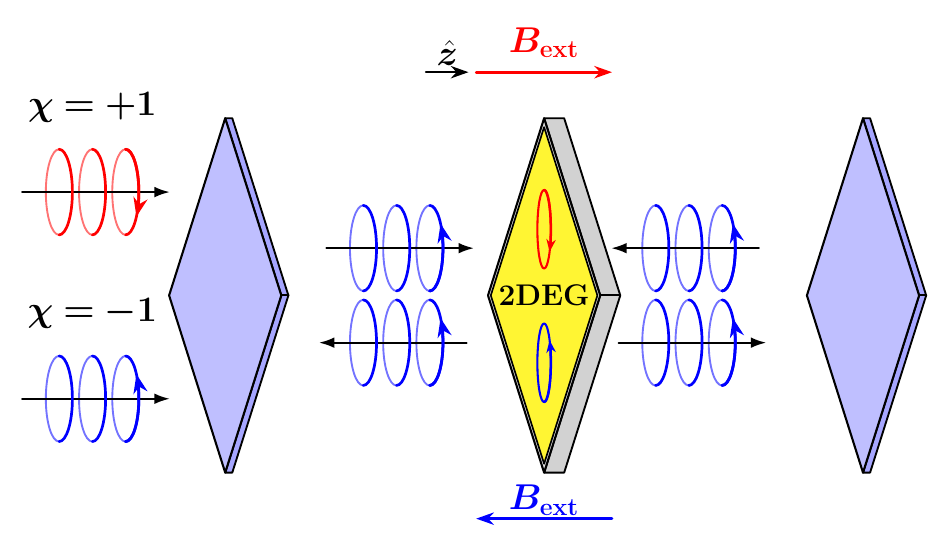}%
}

\end{minipage}%
\hfill%
\begin{minipage}[t]{0.485\textwidth}
\vspace{0pt}

\makebox[0pt][l]{%
    \hspace{\panelBLabelX}%
    \raisebox{\panelBLabelY}[0pt][0pt]{\textbf{b}}%
}

\hspace{\panelBImageX}%
\raisebox{\panelBImageY}{%
\resizebox{\panelBWidth}{!}{%
\begin{tikzpicture}[>=Stealth]

\begin{scope}[shift={(-3.25,0)}]

    \node[
        font=\Large\bfseries,
        align=center,
        text=blue!80!black
    ] at (0,3.85)
    {Resonant Interaction};

    \node[
        font=\Large,
        text=black!70
    ] at (0,3)
    {Co-rotating};

    \filldraw[
        fill=yellow!10,
        draw=yellow!70!orange,
        line width=1.2pt,
        rounded corners=2pt
    ]
    (-2.5,-1.2) -- (1.5,-1.2) --
    (2.5,1.2) -- (-1.5,1.2) -- cycle;

    \node[
        font=\Large\bfseries,
        text=black!90
    ] at (1.12,0.93) {2DEG};

    \draw[
        line width=2pt,
        blue!85!black
    ] (0,0) ellipse (1.6 and 0.5);

    \draw[
        -{Stealth[length=2.5mm]},
        line width=2pt,
        blue!85!black
    ]
    (-0.15,0.50)
    arc[
        start angle=95,
        end angle=25,
        x radius=1.6,
        y radius=0.5
    ];

    \shade[ball color=gray!30]
        (1.6,0) circle (0.15);

    \node[
        font=\LARGE\bfseries,
        right,
        text=gray!70!black
    ] at (1.75,0) {$e^-$};

    \node[
        font=\Large,
        text=blue!85!black
    ] at (-0.3,-0.85)
    {Cyclotron orbit};

    \draw[
        -{Stealth[length=4mm,width=3mm]},
        line width=2.5pt,
        magenta!90!black,
        preaction={
            draw=yellow!10,
            line width=4pt
        }
    ]
    (0,2.2) -- (0,-0.1)
    node[
        pos=0.49,
        left=3pt,
        font=\LARGE,
        text=magenta!90!black
    ] {$\mathbf{B}$};

    \draw[
        line width=2.5pt,
        blue!85!black
    ]
    (0,2.0) ellipse (1.2 and 0.35);

    \draw[
        -{Stealth[length=2.7mm]},
        line width=1.5pt,
        blue!85!black
    ]
    (-0.12,2.35)
    arc[
        start angle=95,
        end angle=35,
        x radius=1.2,
        y radius=0.35
    ];

    \node[
        font=\LARGE\bfseries,
        text=blue!85!black,
        right
    ] at (1.3,2.0) {$a_-$};

\end{scope}

\begin{scope}[shift={(3.25,0)}]

    \node[
        font=\Large\bfseries,
        align=center,
        text=red!85!black
    ] at (0,3.85)
    {Anti-resonant Interaction};

    \node[
        font=\Large,
        text=black!70
    ] at (0,3)
    {Counter-rotating};

    \filldraw[
        fill=yellow!10,
        draw=yellow!70!orange,
        line width=1.2pt,
        rounded corners=2pt
    ]
    (-2.5,-1.2) -- (1.5,-1.2) --
    (2.5,1.2) -- (-1.5,1.2) -- cycle;

    \node[
        font=\Large\bfseries,
        text=black!90
    ] at (1.12,0.93) {2DEG};

    \draw[
        line width=2pt,
        red!80!black
    ]
    (0,0) ellipse (1.6 and 0.5);

    \draw[
        -{Stealth[length=2.5mm]},
        line width=2pt,
        red!80!black
    ]
    (0.15,0.50)
    arc[
        start angle=86.5,
        end angle=160,
        x radius=1.6,
        y radius=0.5
    ];

    \shade[ball color=gray!30]
        (1.6,0) circle (0.15);

    \node[
        font=\LARGE\bfseries,
        right,
        text=gray!70!black
    ] at (1.75,0) {$e^-$};

    \node[
        font=\Large,
        text=red!80!black
    ] at (-0.3,-0.85)
    {Cyclotron orbit};

    \draw[
        -{Stealth[length=4mm,width=3mm]},
        line width=2.5pt,
        magenta!90!black
    ]
    (0,-0.1) -- (0,2.2)
    node[
        pos=0.52,
        left=3pt,
        font=\LARGE,
        text=magenta!90!black
    ] {$\mathbf{B}$};

    \draw[
        line width=2.5pt,
        blue!85!black
    ]
    (0,2.0) ellipse (1.2 and 0.35);

    \draw[
        -{Stealth[length=2.7mm]},
        line width=1.5pt,
        blue!85!black
    ]
    (-0.12,2.35)
    arc[
        start angle=95,
        end angle=35,
        x radius=1.2,
        y radius=0.35
    ];

    \node[
        font=\LARGE\bfseries,
        text=blue!85!black,
        right
    ] at (1.3,2.0) {$a_-$};

\end{scope}

\end{tikzpicture}%
}%
}%

\end{minipage}


\par\vspace{4mm}

\noindent
\begin{minipage}[t]{0.485\textwidth}
\vspace{0pt}

\makebox[0pt][l]{%
    \hspace{\panelCLabelX}%
    \raisebox{\panelCLabelY}[0pt][0pt]{\textbf{c}}%
}

\hspace{\panelCImageX}%
\raisebox{\panelCImageY}{%
    \includegraphics[
        width=\panelCWidth
    ]{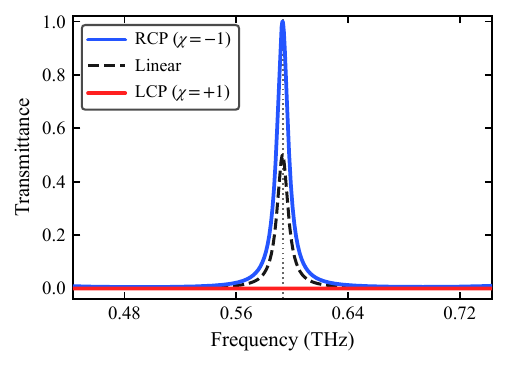}%
}

\end{minipage}%
\hfill%
\begin{minipage}[t]{0.485\textwidth}
\vspace{0pt}

\makebox[0pt][l]{%
    \hspace{\panelDLabelX}%
    \raisebox{\panelDLabelY}[0pt][0pt]{\textbf{d}}%
}

\hspace{\panelDImageX}%
\raisebox{\panelDImageY}{%
    \includegraphics[
        width=\panelDWidth
    ]{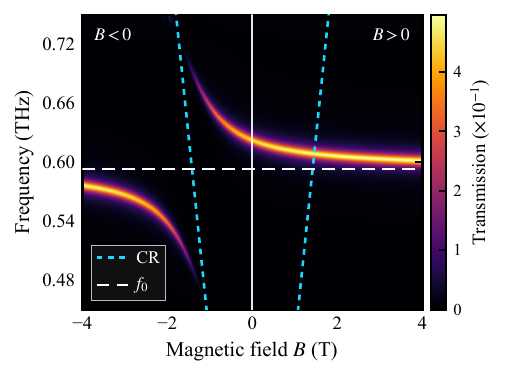}%
}

\end{minipage}



    \caption{
    \textbf{Fixed-polarization cavity coupled to cyclotron resonance.}
    \textbf{a}, Chiral photonic-crystal cavity containing a 2DEG layer.
    The cavity circular polarization is fixed, while reversal of $B$ reverses the
    cyclotron motion.
    \textbf{b}, Interaction scheme for the single-polarization cavity mode $a_-$. 
    For $B<0$, the cyclotron motion rotates in the same sense as $a_-$, producing 
    resonant excitation exchange. Reversing the magnetic field reverses the cyclotron rotation while leaving the cavity mode unchanged, so for $B>0$ the interaction becomes anti-resonant.
    \textbf{c}, Bare cavity transmission spectra without the 2DEG under 
    RCP, LCP and linearly polarized excitation, showing an pronounced
    RCP cavity resonance and strongly suppressed LCP transmission.
    \textbf{d}, Magnetotransmission spectra of the coupled cavity--2DEG system calculated using the transfer-matrix method for linearly polarized incident light. For $B<0$, the RCP cavity mode is co-rotating with the cyclotron motion and develops an avoided crossing. For $B>0$, the same cavity mode is counter-rotating with the cyclotron motion, and the conventional polaritonic anticrossing disappears, leaving only the vacuum Bloch-Siegert resonance shift.
}

    \label{fig:fig1}

\end{figure}

To relate this ideal limit to concrete cavity structure, we consider
a one-dimensional photonic-crystal cavity (PCC) with chiral mirrors whose polarization selectivity is held fixed (Figure~\ref{fig:fig1}a). This design is inspired by a recently demonstrated chiral cavity incorporating InSb layers. The magnetoplasma response of InSb has enabled both the
theoretical proposal and experimental realization of high-quality-factor
chiral terahertz PCCs, in which gyrotropy selectively suppresses
one circular polarization while preserving the opposite circular cavity
mode~\cite{Tay2025Chiral,Kulkarni2026}. In these implementations, however, reversing the applied magnetic field also reverses the cavity’s polarization selectivity, which is not a desired feature for our purpose. Here, we fix the refractive index of the chiral mirrors independently of the magnetic field to focus on the consequences of a single-polarization chiral cavity. Details of the cavity model are provided in Supplementary Note~4.



We label the two circular polarizations by $\chi=\pm1$, with polarization vectors $\mathbf{e}_{\chi}=(\mathbf{e}_x+i\chi\mathbf{e}_y)/\sqrt{2}$ defined in the fixed laboratory frame. Throughout this work, we refer to $\chi=+1$ and $\chi=-1$ as left-circularly polarized (LCP) and right-circularly polarized (RCP), respectively, irrespective of the propagation direction. The cavity supports only a single RCP mode, $a_-$. The interaction between the CR and the cavity mode $a_-$ can be switched between resonant and anti-resonant regimes by reversing the magnetic field, as illustrated in Fig.~\ref{fig:fig1}b. 

To confirm that the cavity model supports only a single circularly polarized mode, we calculate its electromagnetic response using the transfer-matrix method (TMM) and independently verify the results through full-wave simulations (see Supplementary Notes~4 and~5). The empty-cavity spectrum in Fig.~\ref{fig:fig1}c shows a pronounced RCP resonance at $\omega_0/2\pi\simeq0.593~\mathrm{THz}$, while LCP transmission is strongly suppressed. A linearly polarized probe therefore excites the cavity through its RCP component.

Introducing the 2DEG produces the asymmetric magnetotransmission spectrum as expected from the polarization selectivity (Fig.~\ref{fig:fig1}d). For $B<0$, the cavity field rotates with the cyclotron
motion and excitations can be exchanged resonantly between light and matter,
producing upper and lower polariton branches. Reversing $B$ reverses the
electron orbit while leaving the cavity unchanged. The same cavity mode then
couples through pair creation and annihilation rather than resonant excitation
exchange: the avoided crossing disappears and the cavity-like branch acquires
the vacuum Bloch--Siegert shift.

\FloatBarrier


\section*{Hamiltonian description of interaction switching}

The distinction between the effects of resonant and anti-resonant interactions is obscured in an achiral cavity, where both circularly polarized modes coexist. Reversing the magnetic field switches the resonantly coupled mode
from one circular polarization to the other, so resonant polariton
formation persists for both field directions~\cite{Li2018}. Counter-rotating
processes nevertheless remain important for the quantum ground state, where
they generate virtual excitations, squeezing and light--matter correlations
~\cite{CiutiPRB2005,El-Amrani2026}. The diamagnetic interaction provides an
additional subtlety: when both polarization modes are present, the diamagnetic interaction can contain
anomalous photonic terms and contribute to vacuum fluctuations
~\cite{CiutiPRB2005,Li2018}.

For a single circular cavity mode, this contribution simplifies drastically.
The orthogonality properties of the circular basis give
\begin{equation}
    \bm A_{\mathrm{cav}}^{\,2}
    =
    \mathcal A^2(z)
    \left(2a_-^\dagger a_-+1\right),
    \label{eq:A2_single_helicity}
\end{equation}
so the diamagnetic interaction contains no pair-creation terms. It is number
conserving and only renormalizes the cavity frequency. The complete derivation
from the minimal-coupling Hamiltonian is given in Methods.

We denote the positive cyclotron frequency by
$\omega_c=e|B|/m^\ast$, while the sign of $B$ determines the sense of the
cyclotron rotation. Defining the diamagnetically renormalized cavity frequency
as $\widetilde{\omega}_0=\omega_0+D$, the Hamiltonian of the coupled chiral cavity–2DEG system becomes,
up to an additive constant,
\begin{equation}
    H
    =
    \hbar\widetilde{\omega}_0 a_-^\dagger a_-
    +
    \hbar\omega_c b^\dagger b
    +i\hbar\bar g
    \begin{cases}
        \left(
            a_- b^\dagger-a_-^\dagger b
        \right),
        & B<0,
        \\[2mm]
        \left(
            a_-^\dagger b^\dagger-a_- b
        \right),
        & B>0.
    \end{cases}
    \label{eq:single_helicity_H}
\end{equation}
Here $\bar g=g\sqrt{\omega_c/\omega_0}$ and
$D=\bar g^2/\omega_c$. The coupling strength follows from the vacuum field
of the cavity and the TMM mode profile (Supplementary Note 4). For the cavity considered
here, $g/\omega_0\simeq0.23$, well above the conventional
$g/\omega_0\simeq0.1$ scale for the ultrastrong-coupling regime
~\cite{FornDiazRMP2019,KockumNatRevPhys2019}. 
Note that this numerical value sets the magnitude of the spectral
and ground-state effects, but not their separation. 

The Hamiltonian Eq.~\eqref{eq:single_helicity_H} reveals that their separation follows directly from the magnetic field direction, or equivalently the alignment of the cavity and CR rotation directions: for $B<0$ the
light--matter interaction is exactly number conserving, whereas for
$B>0$ it is purely anti-resonant, apart from the number-conserving
diamagnetic renormalization of the cavity frequency. This separation therefore holds at all coupling strengths, including the
deep-strong-coupling regime, within the validity of the single-mode
model. Increasing the coupling enhances the Rabi splitting and vacuum
dressing without changing their assignment to opposite magnetic-field
orientations.
Equation~\eqref{eq:single_helicity_H} describes the ideal limit of circular-polarization-selective
light--matter coupling. In previous chiral light--matter models,
co- and counter-rotating interactions are associated with photonic modes of opposite
circular polarizations~\cite{PhysRevLett.123.133603}. Here, one of these
photonic sectors is removed from the cavity altogether. Consequently,
reversing the cyclotron motion does not transfer the resonant interaction
to another cavity polarization; instead, it changes the interaction
experienced by the same surviving cavity mode.

\begin{figure}[!t]
    \centering


    \begin{minipage}[t]{0.485\textwidth}
        \vspace{0pt}
        \centering
        \panelgraphic[
            width=\linewidth
        ]{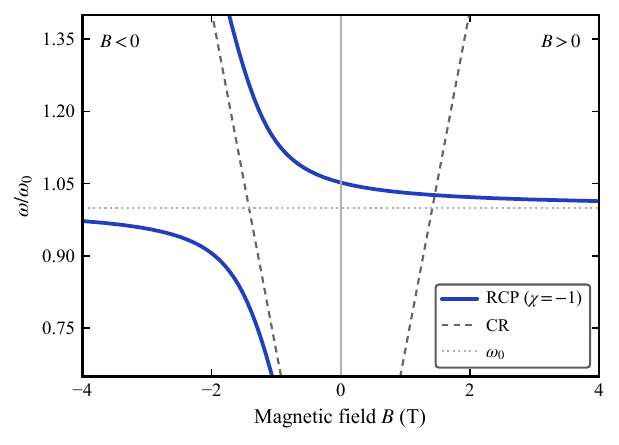}{a}
    \end{minipage}%
    \hfill%
    \begin{minipage}[t]{0.485\textwidth}
        \vspace{0pt}
        \centering
        \panelgraphic[
            width=\linewidth
        ]{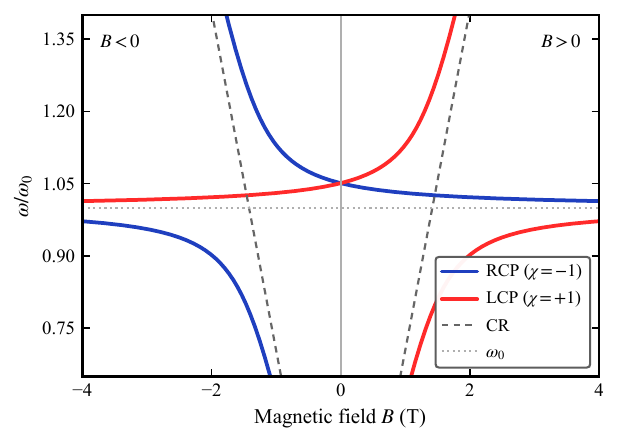}{b}
    \end{minipage}

    \caption{
    \textbf{Polariton spectra in chiral and achiral cavities.}
    \textbf{a}, Polariton spectra in a single-polarization chiral cavity. For $B<0$, the cavity field
    and cyclotron motion rotate in the same sense and form upper and lower
    polaritons. For $B>0$, the cyclotron motion reverses, leading to an anti-resonant interaction with the cavity mode, and the conventional Rabi splitting disappears.
    \textbf{b}, Polariton spectra in a conventional achiral cavity supporting both RCP ($\chi=-1$) and
    LCP ($\chi=+1$) modes, calculated with the same coupling strength $g$.
    For either magnetic field direction, one circular mode participates in
    resonant polariton formation while the other provides the anomalous
    interaction channel.
    }
    \label{fig:helicity_spectra}
\end{figure}

Figure~\ref{fig:helicity_spectra} makes the role of cavity polarization
explicit. In the chiral cavity (Fig.~\ref{fig:helicity_spectra}a), field
reversal changes the type of interaction experienced by the same cavity mode.
The resulting eigenfrequencies closely follow the magnetotransmission of
Fig.~\ref{fig:fig1}d: the polaritonic branches appear only when the cavity
field and cyclotron motion have the same handedness, whereas the opposite
field direction retains the shifted cavity-like branch without a conventional
anticrossing.

A conventional achiral cavity serve as the reference case
(Fig.~\ref{fig:helicity_spectra}b), with its Hamiltonian given in Supplementary Note~2. Because both circular modes are present,
reversing the cyclotron motion simply changes which one participates in
resonant hybridization. The difference between the two spectra is therefore
set by the electromagnetic modes available to the CR, rather than by a
change in coupling strength.

\FloatBarrier


\section*{Separation of vacuum dressing and Rabi splitting}

\begin{figure}[!t]
    \centering


    \begin{minipage}[t]{0.49\linewidth}
        \centering
        \begin{tikzpicture}
            \node[
                inner sep=0,
                anchor=south west
            ] (img) at (0,0) {%
                \includegraphics[
                    width=\linewidth
                ]{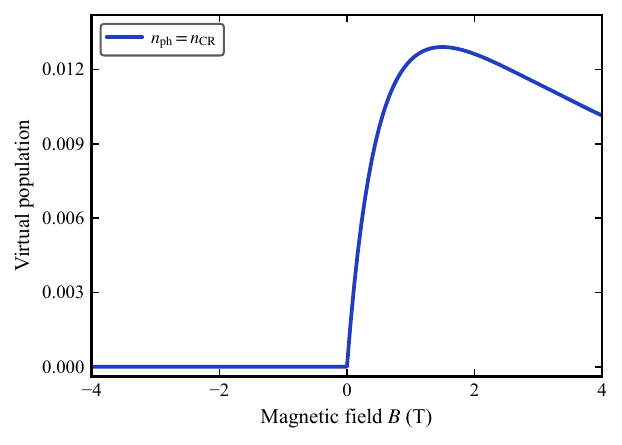}%
            };

            \node[
                font=\bfseries,
                anchor=north west
            ] at ([xshift=1mm,yshift=-1mm]img.north west) {a};
        \end{tikzpicture}
    \end{minipage}
    \hfill
    \begin{minipage}[t]{0.49\linewidth}
        \centering
        \begin{tikzpicture}
            \node[
                inner sep=0,
                anchor=south west
            ] (img) at (0,0) {%
                \includegraphics[
                    width=\linewidth
                ]{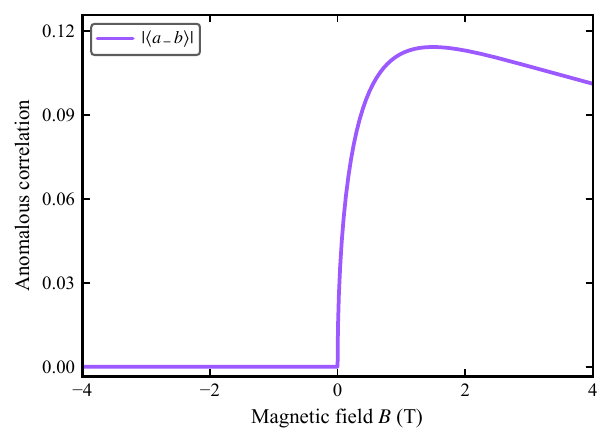}%
            };

            \node[
                font=\bfseries,
                anchor=north west
            ] at ([xshift=1mm,yshift=-1mm]img.north west) {b};
        \end{tikzpicture}
    \end{minipage}

    \vspace{3mm}


    \begin{minipage}[t]{0.49\linewidth}
        \centering
        \begin{tikzpicture}
            \node[
                inner sep=0,
                anchor=south west
            ] (img) at (0,0) {%
                \includegraphics[
                    width=\linewidth
                ]{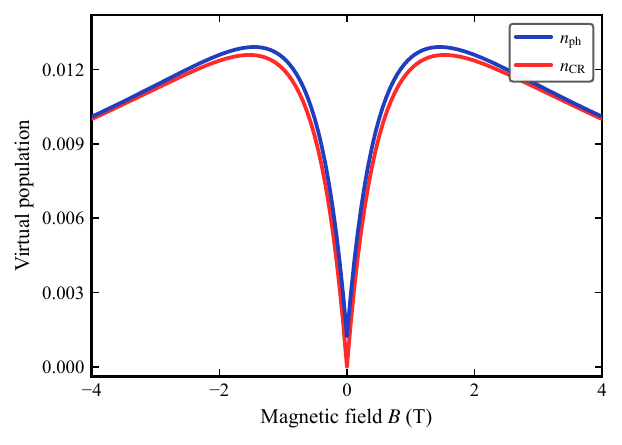}%
            };

            \node[
                font=\bfseries,
                anchor=north west
            ] at ([xshift=1mm,yshift=-1mm]img.north west) {c};
        \end{tikzpicture}
    \end{minipage}
    \hfill
    \begin{minipage}[t]{0.49\linewidth}
        \centering
        \begin{tikzpicture}
            \node[
                inner sep=0,
                anchor=south west
            ] (img) at (0,0) {%
                \includegraphics[
                    width=\linewidth
                ]{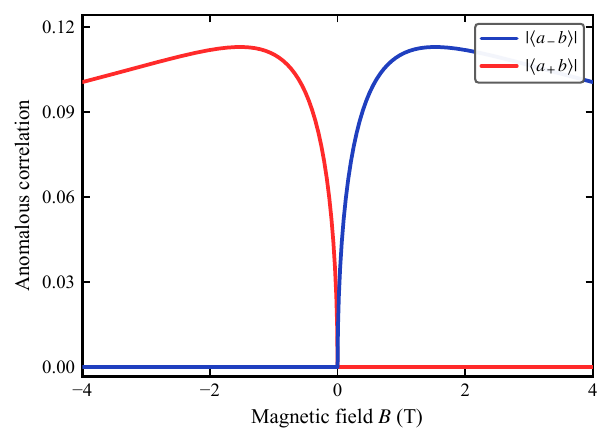}%
            };

            \node[
                font=\bfseries,
                anchor=north west
            ] at ([xshift=1mm,yshift=-1mm]img.north west) {d};
        \end{tikzpicture}
    \end{minipage}

    \caption{
    \textbf{Selective emergence of quantum vacuum dressing in a chiral cavity.}
    \textbf{a}, Virtual photon and CR populations in the single-polarization chiral cavity. They vanish for $B<0$ and become equal and finite for $B>0$.
    \textbf{b}, Corresponding anomalous photon--CR correlation the chiral cavity.
    \textbf{c}, Total virtual photon and CR populations in the conventional achiral
    cavity.
    \textbf{d}, Anomalous correlations between the CR and the two circular
    cavity modes in the achiral cavity.
    }
    \label{fig:vacuum_observables}
\end{figure}

The distinction is sharper in the ground state, which we characterize through virtual populations and the anomalous correlation $|\langle a_-b\rangle|$ (Figs.~\ref{fig:vacuum_observables}a,b). For the chiral cavity at $B<0$, the
complete Hamiltonian conserves excitation number, including the diamagnetic
term, and the bare vacuum remains its exact ground state despite the large
Rabi splitting. The virtual photon population, CR population and anomalous
correlation therefore all vanish. For $B>0$, light and matter excitations are instead created and annihilated
in pairs. In the absence of conventional Rabi splitting, the ground state becomes a correlated vacuum with finite
$n_{\mathrm{ph}}=n_{\mathrm{CR}}$ and
$|\langle a_-b\rangle|$, both of which peak near the zero-detuning magnetic field.

The coexistence of ultrastrong coupling and a trivial ground state has also been reported in purely co-rotating magnonic Hopfield systems \cite{Wang2024}. Here, however, the complementary bare and dressed vacuum regimes are accessed within the same cavity--matter system by reversing only the magnetic field. In a different context,
Stokes and Nazir showed that Jaynes--Cummings physics can persist in
the ultrastrong-coupling regime when finite-level matter truncation is
treated consistently with the choice of gauge
\cite{StokesNazirNatCommun2019}. The present mechanism is distinct, as no matter-level truncation is made.
Starting from the minimal-coupling Hamiltonian of the 2DEG, we retain
the full bosonic cyclotron degree of freedom together with the
diamagnetic interaction. The number-conserving interaction for $B<0$
therefore follows from circular-polarization selection rather than from a
truncated matter description or a particular gauge choice.


The achiral cavity remains vacuum dressed for either field direction (Figs.~\ref{fig:vacuum_observables}c,d);
For the same normalized coupling strength $g/\omega_0$, its
total virtual photon population is nearly the same as in the dressed regime of the single-polarization
case, whereas its CR population is slightly reduced. The additional cavity
mode therefore redistributes part of the photonic vacuum dressing from $a_-$ to $a_+$ rather than simply adding more virtual excitations.
The anomalous light--matter correlation remains concentrated between the CR and the counter-rotating cavity mode, while the correlation with the co-rotating cavity mode is negligible. This redistribution reveals a role of the diamagnetic interaction beyond its
familiar spectral blue shift. When both circular polarizations are available, its
anomalous contribution provides an additional route for redistributing vacuum fluctuations within the cavity field.

\section*{Chiral vacuum-field fluctuations}

The chirality dependence of the ground state is encoded directly in the
electromagnetic vacuum fluctuations. We can characterize them through the
symmetrized electric-field correlation function
\begin{equation}
    G_E^{(\pm)}(z,z';\tau)
    =
    \frac{1}{2}
    \left\langle
        \left\{
            \delta\hat E(z,\tau),
            \delta\hat E(z',0)
        \right\}
    \right\rangle_{\pm B},
    \label{eq:field_correlation}
\end{equation}
where the two signs denote opposite magnetic-field orientations at the same
$|B|$. For $B<0$, the Hamiltonian is number conserving and its ground state
is the bare vacuum. For $B>0$, the anti-resonant interaction generates a
two-mode squeezed ground state with a finite virtual photon population
$n_{\rm ph}=\langle a_-^\dagger a_-\rangle$. In the present
single-polarization model, $\langle a_-a_-\rangle=0$, and the local
electric-field fluctuations therefore satisfy
\begin{equation}
    G_E^{(-)}(z,z;0)
    =
    |E_{\rm vac}(z)|^2,
    \qquad
    G_E^{(+)}(z,z;0)
    =
    \left(1+2n_{\rm ph}\right)
    |E_{\rm vac}(z)|^2 .
    \label{eq:field_variance}
\end{equation}
Consequently,
\begin{equation}
    \frac{
        G_E^{(+)}(z,z;0)
        -
        G_E^{(-)}(z,z;0)
    }{
        G_E^{(-)}(z,z;0)
    }
    =
    2n_{\rm ph}.
    \label{eq:field_variance_contrast}
\end{equation}
Thus, even the electric-field fluctuations inherit the separation by the cavity chirality: the resonant regime can exhibit a large polaritonic splitting
without excess ground-state field noise, whereas reversing the field produces
an excess vacuum fluctuation despite the disappearance of the conventional
anticrossing.

Electric-field vacuum fluctuations and their temporal and spatial correlations
have been accessed experimentally in the terahertz regime using ultrafast
electro-optic sampling~\cite{Riek2015Vacuum,BeneaChelmus2019Vacuum,
Settembrini2022Vacuum}. In particular, electro-optic detection provides access
to field correlations rather than relying on photon emission from the ground
state. An analogous differential measurement under magnetic-field reversal would therefore
provide a possible route to probing the chiral vacuum contrast predicted here.
Within the present model, the normalized field-reversal differential signal in
Eq.~\eqref{eq:field_variance_contrast} directly tracks the virtual photon
population. Note that this does not imply direct detection
of virtual photons as freely propagating radiation: ground-state virtual
excitations are bound to the interacting system, and converting them into
real photons generally requires an additional nonadiabatic or nonequilibrium
process~\cite{DeLiberato2017Virtual}.

\FloatBarrier


\section*{From vacuum dressing to matter engineering}

An important question raised by this separation is which properties of the
quantum vacuum are actually responsible for modifying matter. Two-dimensional electron gases are particularly suited to this question,
since cavity vacuum fields have already been shown to modify their
magnetotransport and quantum Hall properties
~\cite{ParaviciniBagliani2019,Appugliese2022,Enkner2025,Graziotto2026}. Recent work
has identified ground-state light--matter entanglement as a key ingredient
in cavity-induced changes of electronic properties~\cite{Passetti2023}. In
the present system, the same pair-creation processes that dress the vacuum
also generate squeezing and light--matter entanglement, whereas the regime
with a large Rabi splitting retains an unentangled bare ground state.
Cavity chirality therefore provides a setting in which the effects of
vacuum entanglement and polariton formation on matter can be tested independently.
For each magnetic-field orientation, comparing a
matter observable in the chiral cavity-coupled system with its corresponding uncoupled reference could reveal whether the cavity-induced change follows spectral hybridization or anomalous vacuum correlations.

The separation principle enabled by a chiral cavity could be extended from selectively controlling polariton formation and vacuum dressing to
controlling where vacuum correlations and entanglement reside within a multimode light--matter system.
Multimode Landau-polariton systems already provide a natural setting for
such an extension. Coupling a common cavity field to several
magnetoplasmon excitations has enabled cooperative multimode
deep-strong coupling and vacuum-mediated entanglement between otherwise
orthogonal electronic modes
~\cite{Mornhinweg2024ModeMultiplexing}. More recently, simultaneous
coupling of the zero-momentum cyclotron resonance and finite-momentum
magnetoplasmons to the same cavity mode has been demonstrated
experimentally, establishing direct cavity-mediated hybridization between
local and nonlocal matter excitations
~\cite{Endo2025CavityMediated}. In such a multimode setting, polarization-selective vacuum interactions could offer a route to controlling how anomalous correlations are distributed among matter modes, exploiting the chiral structure of the vacuum itself~\cite{El-Amrani2026}.

A complementary route would be to engineer a cavity supporting
non-degenerate modes of opposite circular polarization. Separately tunable mode frequencies would provide
independent photonic handles on the resonant and anti-resonant interaction
channels, allowing spectral hybridization, vacuum dressing and the
associated quantum correlations to be tuned with substantially greater
freedom than in a conventional achiral cavity. Combining such
polarization-selective photonic modes with a multimode matter sector could
therefore provide finer control over both the strength and the
structure of vacuum-mediated correlations.

The underlying polarization-selection principle is not unique to cyclotron resonance.
In microwave cavity magnonics, where magnetization precession also carries
a definite handedness, circular cavity fields have been used to select the
coupling to one sense of rotation, while reversal of the magnetic bias can
switch between level repulsion and level crossing
~\cite{Bourhill2023,Joseph2024}. Related separations of resonant and
anti-resonant interactions have also been explored in anisotropic magnonic
Hopfield systems, where the two coupling channels can be varied
independently and either one can become dominant
~\cite{Makihara2021,Wang2024}. In the present setting, the analogous
control arises from the frequency and polarization of the cavity mode rather than from magnetic anisotropy.

Several existing photonic architectures offer tools for
engineering chiral cavity modes. Chiral subwavelength terahertz cavities and chiral
photonic-crystal cavities demonstrate polarization-selective modes in the
Landau-polariton and magneto-optical regimes
~\cite{PhysRevB.109.L161302,Tay2025Chiral,Kulkarni2026}. In these
implementations, however, cavity chirality remains tied to a magnetic
bias or to the light--matter interaction itself, limiting independent control over cavity chirality and the matter modes. Spin-preserving
photonic-crystal mirrors, chiral metamirrors and metasurface cavities
provide further means of engineering the polarization properties of photonic environments
~\cite{Semnani2020SpinMirror,Plum2015ChiralMirror,
Bassler2024MetasurfaceCavity,Yan2025THzMirror}. Although these approaches
do not directly realize the fixed-chirality cavity considered here, they
offer building blocks that could aid the development of cavities whose chirality is controlled independently of the magnetic field applied to the 2DEG.

Our results establish cavity chirality as a control parameter with distinct effects on the excitation spectrum and on the interacting
quantum vacuum. A large Rabi splitting and a strongly dressed vacuum are
separable consequences of ultrastrong coupling: they can
be selected independently within the same light--matter system. Extending
this principle to non-degenerate chiral cavity modes and multimode matter
systems could enable polariton formation, vacuum
dressing, and the distribution of quantum correlations to be controlled as complementary
resources for engineering cavity quantum materials.

\FloatBarrier

\section*{Note added}
During the final preparation of this manuscript, we became aware of a related independent preprint that became publicly available on 5 October 2026~\cite{Kulkarni2026ChiralVacuum}

\section*{Methods}

\subsection*{Transfer matrix calculation}

The magnetotransmission is calculated using a one-dimensional transfer matrix
method at normal incidence. The two circular components propagate
independently, while the optical response of the chiral mirrors is held fixed
throughout the magnetic field sweep. The detailed cavity structure and mirror
parametrization are given in Supplementary Note~4.

The magnetic field dependence enters through the 2DEG. In the circular basis,
its dielectric response is written as
\begin{equation}
    \epsilon_{\mathrm{CRA}}(\omega,B)
    =
    \epsilon_{\mathrm{bg}}
    -
    \frac{\omega_p^2}
    {\omega\left(\omega-\omega_c+i\gamma\right)},
\end{equation}
and
\begin{equation}
    \epsilon_{\mathrm{CRI}}(\omega,B)
    =
    \epsilon_{\mathrm{bg}}
    -
    \frac{\omega_p^2}
    {\omega\left(\omega+\omega_c+i\gamma\right)},
\end{equation}
where
\begin{equation}
    \omega_c=\frac{e|B|}{m^\ast}
\end{equation}
is the positive cyclotron frequency and $\gamma$ is the cyclotron linewidth.
Reversing $B$ exchanges which circular component is co-rotating with the
cyclotron motion while leaving the cavity response unchanged. For a linearly
polarized probe,
\begin{equation}
    T_{\mathrm{lin}}
    =
    \frac{1}{2}
    \left(
        T_{\mathrm{RCP}}+T_{\mathrm{LCP}}
    \right).
\end{equation}
The numerical parameters and complete transfer matrix implementation are
provided in Supplementary Note~4.

\subsection*{Minimal coupling Hamiltonian}

We describe the 2DEG within the parabolic-band approximation and start from
the minimal-coupling Hamiltonian
\begin{equation}
    H
    = H_{\mathrm{cav}}+
    \sum_j
    \frac{
    \left[
        \bm p_j
        +e\bm A_{\mathrm{ext}}(\bm r_j)
        +e\bm A_{\mathrm{cav}}(\bm r_j)
    \right]^2
    }{2m^\ast},
\end{equation}
where $H_{\mathrm{cav}}$ is the cavity Hamiltonian, $\bm A_{\mathrm{ext}}$ generates the static magnetic field and
$\bm A_{\mathrm{cav}}$ is the quantized cavity field. The spatially uniform
in-plane current couples to the collective cyclotron mode, represented by the
bosonic operator $b$. We take $\omega_c=e|B|/m^\ast>0$, while the sign of
$B$ fixes the sense of the cyclotron rotation.

For the single-polarization cavity considered in the main text, the supported
mode is RCP, $\chi=-1$, and the Hermitian cavity field has the form
\begin{equation}
    \bm A_{\mathrm{cav}}(z)
    =
    \mathcal A(z)
    \left(
        \bm e_- a_-
        +
        \bm e_+ a_-^\dagger
    \right),
\end{equation}
with
\begin{equation}
    \bm e_\chi
    =
    \frac{\bm e_x+i\chi\bm e_y}{\sqrt{2}} .
\end{equation}
Because
$\bm e_\chi\!\cdot\!\bm e_\chi=0$ and
$\bm e_\chi\!\cdot\!\bm e_{-\chi}=1$,
the diamagnetic contribution reduces to
\begin{equation}
    \bm A_{\mathrm{cav}}^{\,2}
    =
    \mathcal A^2(z)
    \left(
        2a_-^\dagger a_-+1
    \right).
\end{equation}
It therefore contains no anomalous photonic terms and only renormalizes the
cavity frequency. Retaining both the paramagnetic and diamagnetic
contributions gives, up to an additive constant,
\begin{equation}
    H_{\mathrm{chiral}}
    =
    \hbar\widetilde{\omega}_0 a_-^\dagger a_-
    +
    \hbar\omega_c b^\dagger b
    +
    i\hbar\bar g
    \begin{cases}
        a_-b^\dagger-a_-^\dagger b,
        & B<0,
        \\[1mm]
        a_-^\dagger b^\dagger-a_-b,
        & B>0 ,
    \end{cases}
    \label{eq:H_1h_methods}
\end{equation}
where
\begin{equation}
    \widetilde{\omega}_0=\omega_0+D,
    \qquad
    D=\frac{\bar g^2}{\omega_c},
    \qquad
    \bar g=g\sqrt{\frac{\omega_c}{\omega_0}} .
\end{equation}
Thus field reversal changes the same cavity mode from resonant excitation
exchange to anomalous pair creation without changing the magnitude of the
coupling.

For comparison, the non-chiral reference retains two degenerate circular
cavity modes $a_+$ and $a_-$. Its vector potential contains both circular modes,
so that the diamagnetic term also generates an anomalous coupling between the
two circular sectors. The resulting Hamiltonian is
\begin{align}
    H_{\mathrm{achiral}}
    ={}&
    \hbar\widetilde{\omega}_0
    \left(
        a_+^\dagger a_+
        +
        a_-^\dagger a_-
    \right)
    +
    \hbar\omega_c b^\dagger b
    \nonumber\\
    &+
    \hbar D
    \left(
        a_+a_-
        +
        a_+^\dagger a_-^\dagger
    \right)
    \nonumber\\
    &+
    i\hbar\bar g
    \begin{cases}
        a_-b^\dagger-a_-^\dagger b
        +a_+^\dagger b^\dagger-a_+b,
        & B<0,
        \\[1mm]
        a_+b^\dagger-a_+^\dagger b
        +a_-^\dagger b^\dagger-a_-b,
        & B>0 .
    \end{cases}
    \label{eq:H_2h_methods}
\end{align}
This circular decomposition is the same physical separation between the
cyclotron-active and cyclotron-inactive sectors underlying the
polarization-resolved Landau-polariton response and vacuum Bloch--Siegert
shift observed in Ref.~\cite{Li2018}. Reversing $B$ simply exchanges the roles
of $a_+$ and $a_-$, so a resonant excitation-exchange channel remains present
for either field direction.

The complete quantization and algebra leading to
Eqs.~\eqref{eq:H_1h_methods} are given in
Supplementary Note~2.

\subsection*{Ground state observables}

Both Hamiltonians are quadratic in bosonic operators. The chiral-cavity
Hamiltonian for $B<0$ is number conserving and is diagonalized by a unitary
rotation of $a_-$ and $b$; its ground state is therefore the bare vacuum.
For $B>0$, the anomalous interaction requires a bosonic Bogoliubov
transformation and produces a correlated two-mode ground state.

The Hamiltonian for a conventional achiral cavity is diagonalized by a Hopfield--Bogoliubov
transformation including both circular cavity sectors. The ground state
covariance matrix is then used to evaluate the virtual populations and
anomalous correlations shown in Fig.~\ref{fig:vacuum_observables}. Details
of the paraunitary diagonalization are given in Supplementary Note~3.

\section*{Data availability}

The data supporting the findings of this study are available in the associated GitHub repository at \url{https://github.com/MoorishQubit/helicity/}.

\section*{Code availability}

The codes used to perform the calculations, simulations and data analysis reported in this study are available in the associated GitHub repository at \url{https://github.com/MoorishQubit/helicity/}.

\section*{Acknowledgements}

A.E-A. thanks Jean Decker for the helpful discussions. Z.M. acknowledges funding from the Ministry of Economic Affairs, Labour and Tourism Baden-Württemberg in the frame of the Competence Center Quantum Computing Baden-Württemberg (project ``KQCBW25'').
H.R.’s material is based upon work supported by the Department of Energy [Basic Energy Sciences] under Award Number(s) [DE-FG02-02ER15344].

\section*{Author contributions}

A.E-A. and D.L. conceived the research. A.E-A. developed the theoretical model, performed the transfer matrix calculations, carried out the numerical analysis and analysed the results. N.B. performed the COMSOL simulations and validation. Z.M. and H.S. contributed to the analytical derivations and the interpretation of the results. H.R., A.E.F. and D.L. supervised the research and provided scientific guidance and feedback. A.E-A. and D.L. wrote the manuscript with input from all authors.

\section*{Competing interests}

The authors declare no competing interests.

\nolinenumbers
\renewcommand{\refname}{References}
\putbib[bibliography]
\end{bibunit}

\clearpage

\resetauthblk
\setcounter{footnote}{0}
\setcounter{equation}{0}
\setcounter{figure}{0}
\setcounter{table}{0}
\setcounter{linenumber}{1}

\renewcommand{\theequation}{S\arabic{equation}}
\renewcommand{\thefigure}{S\arabic{figure}}
\renewcommand{\thetable}{S\arabic{table}}

\renewcommand{\theHequation}{S\arabic{equation}}
\renewcommand{\theHfigure}{S\arabic{figure}}
\renewcommand{\theHtable}{S\arabic{table}}

\setcounter{page}{1}
\renewcommand{\thepage}{S\arabic{page}}

\begin{bibunit}

\begin{center}

{\Large\bfseries Supplementary Information for\par}
\vspace{0.5em}

{\LARGE
Decoupling ultrastrong Rabi splitting from quantum
vacuum dressing in a chiral cavity
\par}

\vspace{1.2em}

{\normalsize
Ayoub El-Amrani\textsuperscript{1,*},
Nasir Boubekraoui\textsuperscript{1},
Zakaria Mzaouali\textsuperscript{2,3},
Houssam Sabri\textsuperscript{1},
Herschel Rabitz\textsuperscript{4},
Abdelouahed El Fatimy\textsuperscript{1},
and Dukhyung Lee\textsuperscript{1,*}
\par}

\vspace{0.8em}

{\small
\textsuperscript{1}College of Physical Sciences and Engineering,
University Mohammed VI Polytechnic, Ben Guerir 43150, Morocco\par

\textsuperscript{2}Institut f\"ur Theoretische Physik,
Universit\"at T\"ubingen, Auf der Morgenstelle 14,
72076 T\"ubingen, Germany\par

\textsuperscript{3}J\"ulich Supercomputing Centre,
Forschungszentrum J\"ulich, J\"ulich 52428, Germany\par

\textsuperscript{4}Department of Chemistry,
Princeton University, Princeton, New Jersey 08544, USA\par
}

\vspace{0.7em}

{\small
\textsuperscript{*}Correspondence:
\texttt{ayoub.el-amrani@um6p.ma};
\texttt{dukhyung.lee@um6p.ma}
\par}

\end{center}

\maketitle

\setcounter{figure}{0}
\renewcommand{\thefigure}{S\arabic{figure}}


\section*{Supplementary Note 1: Quantization of the chiral cavity field}

We first quantize the electromagnetic field of the chiral cavity. We start
from the classical cavity Hamiltonian
\begin{equation}
    H_{\mathrm{cav}}
    =
    \int d\bm r
    \left[
        \frac{\epsilon_0}{2}
        \bm E_{\perp}^{\,2}(\bm r)
        +
        \frac{1}{2\mu_0}
        \bm B^{2}(\bm r)
    \right],
    \label{eq:Hclassical}
\end{equation}
where $\bm E_{\perp}(\bm r)$ is the transverse electric field and
$\bm B(\bm r)$ is the magnetic field. It is convenient to work in terms of
the vector potential $\bm A(\bm r)$ and its canonical conjugate momentum
$\bm\Pi(\bm r)$, defined through
\begin{equation}
    \bm B(\bm r)
    =
    \bm\nabla\times\bm A(\bm r),
    \qquad
    \bm E_{\perp}(\bm r)
    =
    -\frac{\bm\Pi(\bm r)}{\epsilon_0},
\end{equation}
with the canonical commutation relation
\begin{equation}
    \left[
        A_i(\bm r),
        \Pi_j(\bm r')
    \right]
    =
    i\hbar\,
    \delta^{\perp}_{ij}(\bm r-\bm r'),
\end{equation}
where $\bm\delta_\perp$ denotes the transverse delta-function tensor.

Before quantization we introduce the electromagnetic approximations relevant
to the present cavity. The response of the dielectric environment, including
the non-resonant background of the two-dimensional electron gas (2DEG), is
described by an effective dielectric function. We consider a
one-dimensional photonic-crystal cavity, so that this response depends only
on the longitudinal coordinate $z$. Chirality is incorporated by allowing
the electromagnetic eigenproblem to depend on the circular polarization index
$\chi=\pm1$, through the dielectric functions
$\epsilon_\chi(z)$.

Throughout this work we define the circular basis as
\begin{equation}
    \bm e_\chi
    =
    \frac{
        \bm e_x+i\chi\bm e_y
    }{\sqrt{2}},
    \qquad
    \chi=\pm1 .
    \label{eq:circular_basis_definition}
\end{equation}
For propagation along $+z$, our convention is
\begin{equation}
    \boxed{
    \chi=-1\equiv\mathrm{RCP},
    \qquad
    \chi=+1\equiv\mathrm{LCP}.
    }
    \label{eq:helicity_convention}
\end{equation}
Thus, $\chi$ is the circular-polarization index used throughout the main text and
Supplementary Information.

The effective cavity Hamiltonian can then be decomposed into the two sectors as:
\begin{equation}
    H_{\mathrm{cav}}
    =
    S
    \sum_{\chi=\pm1}
    \int dz
    \left[
        \frac{
            \bm\Pi_\chi^{\,2}(z)
        }{
            2\epsilon_0\epsilon_\chi(z)
        }
        +
        \frac{1}{2\mu_0}
        \left(
            \frac{\partial\bm A_\chi(z)}{\partial z}
        \right)^2
    \right],
    \label{eq:Hcav}
\end{equation}
where $S$ is the transverse quantization area.

The index $\chi$ in Eq.~\eqref{eq:Hcav} labels independent dynamical
normal-mode sectors of definite circular polarization. It should not be confused with a
simple projection of the total Hermitian field along $\bm e_\chi$. In
particular, a Hermitian field associated with a circularly polarized mode
contains both its positive-frequency component and the Hermitian conjugate
negative-frequency component. This distinction becomes explicit below.

Hamilton's equations are
\begin{equation}
    \frac{\partial}{\partial t}
    \bm A_\chi(z,t)
    =
    \frac{\delta H_{\mathrm{cav}}}
    {\delta\bm\Pi_\chi(z,t)}
    =
    \frac{
        \bm\Pi_\chi(z,t)
    }{
        \epsilon_0\epsilon_\chi(z)
    },
    \label{eq:Hamilton_A}
\end{equation}
and
\begin{equation}
    \frac{\partial}{\partial t}
    \bm\Pi_\chi(z,t)
    =
    -
    \frac{\delta H_{\mathrm{cav}}}
    {\delta\bm A_\chi(z,t)}
    =
    \frac{1}{\mu_0}
    \frac{\partial^2}{\partial z^2}
    \bm A_\chi(z,t).
    \label{eq:HamiltonPi}
\end{equation}
Combining Eqs.~\eqref{eq:Hamilton_A} and \eqref{eq:HamiltonPi} gives
\begin{equation}
    \frac{\partial^2}{\partial z^2}
    \bm A_\chi(z,t)
    -
    \frac{\epsilon_\chi(z)}{c^2}
    \frac{\partial^2}{\partial t^2}
    \bm A_\chi(z,t)
    =
    0 .
    \label{eq:waveequation}
\end{equation}

We first expand the positive-frequency part of the field in normal modes of
circular polarization $\chi$,
\begin{equation}
    \bm A_\chi^{(+)}(z,t)
    =
    \sum_n
    \frac{
        f_{n,\chi}(z)
    }{
        \sqrt{\epsilon_\chi(z)}
    }
    \alpha_{n,\chi}
    e^{-i\omega_{n,\chi}t}
    \bm e_\chi .
    \label{eq:Aexpansion}
\end{equation}
Substituting Eq.~\eqref{eq:Aexpansion} into
Eq.~\eqref{eq:waveequation} gives
\begin{equation}
    -
    \frac{1}{\sqrt{\epsilon_\chi(z)}}
    \frac{d^2}{dz^2}
    \left[
        \frac{
            f_{n,\chi}(z)
        }{
            \sqrt{\epsilon_\chi(z)}
        }
    \right]
    \equiv
    \mathcal L_\chi f_{n,\chi}(z)
    =
    \frac{
        \omega_{n,\chi}^2
    }{
        c^2
    }
    f_{n,\chi}(z).
    \label{eq:hermitian_wave_equation}
\end{equation}

From this point onward we assume that the circular-polarization-dependent dielectric
functions $\epsilon_\chi(z)$ are real. This defines the ideal lossless
chiral cavity that is canonically quantized here. Realistic implementations
contain radiative leakage and absorption, but the conservative normal-mode
problem provides the appropriate starting point for the closed-system
Hamiltonian. Dissipation can subsequently be incorporated through linewidths
in linear response or, more formally, by coupling the cavity modes to
external bath degrees of freedom.

Under this assumption, Eq.~\eqref{eq:hermitian_wave_equation} defines a
self-adjoint eigenvalue problem in each circular-polarization sector. The mode functions
$f_{n,\chi}(z)$ can therefore be chosen real and orthonormal:
\begin{equation}
    \sum_n
    f_{n,\chi}(z)
    f_{n,\chi}(z')
    =
    \delta(z-z'),
    \label{eq:f_completeness}
\end{equation}
and
\begin{equation}
    \int dz\,
    f_{n,\chi}(z)
    f_{n',\chi}(z)
    =
    \delta_{nn'} .
    \label{eq:f_orthonormality}
\end{equation}

In an ordinary non-chiral cavity, the two circular-polarization sectors may be degenerate.
In a chiral cavity they need not be: their eigenfrequencies and spatial
profiles can differ, and in an open cavity their linewidths may differ as
well.

Equation~\eqref{eq:Aexpansion} contains only the positive-frequency part.
The corresponding Hermitian vector potential is therefore
\begin{align}
    \bm A_\chi(z,t)
    =
    \sum_n
    \frac{
        f_{n,\chi}(z)
    }{
        \sqrt{\epsilon_\chi(z)}
    }
    \Big[
        &
        \alpha_{n,\chi}
        e^{-i\omega_{n,\chi}t}
        \bm e_\chi
        \nonumber\\
        &+
        \alpha_{n,\chi}^{*}
        e^{i\omega_{n,\chi}t}
        \bm e_\chi^{*}
    \Big].
    \label{eq:realA}
\end{align}
This expression makes clear that $\bm A_\chi$ is a Hermitian field belonging
to the circular-polarization sector labeled by $\chi$, rather than simply the scalar
projection of the total field onto $\bm e_\chi$.

Using Eq.~\eqref{eq:Hamilton_A}, the corresponding canonical momentum is
\begin{align}
    \bm\Pi_\chi(z,t)
    =
    -i\epsilon_0
    \sum_n
    \sqrt{\epsilon_\chi(z)}
    f_{n,\chi}(z)
    \omega_{n,\chi}
    \Big[
        &
        \alpha_{n,\chi}
        e^{-i\omega_{n,\chi}t}
        \bm e_\chi
        \nonumber\\
        &-
        \alpha_{n,\chi}^{*}
        e^{i\omega_{n,\chi}t}
        \bm e_\chi^{*}
    \Big].
    \label{eq:Pi_expansion}
\end{align}

We now substitute Eqs.~\eqref{eq:realA} and
\eqref{eq:Pi_expansion} back into Eq.~\eqref{eq:Hcav}. Suppressing the
explicit time dependence, the canonical-momentum contribution is
\begin{align}
    &
    S
    \int dz\,
    \frac{
        \bm\Pi_\chi^{\,2}(z)
    }{
        2\epsilon_0\epsilon_\chi(z)
    }
    \nonumber\\
    &=
    -
    \frac{S\epsilon_0}{2}
    \sum_{n,m}
    \omega_{n,\chi}
    \omega_{m,\chi}
    \int dz\,
    f_{n,\chi}(z)
    f_{m,\chi}(z)
    \nonumber\\
    &\qquad\times
    \left(
        \alpha_{n,\chi}\bm e_\chi
        -
        \alpha_{n,\chi}^{*}\bm e_\chi^{*}
    \right)
    \cdot
    \left(
        \alpha_{m,\chi}\bm e_\chi
        -
        \alpha_{m,\chi}^{*}\bm e_\chi^{*}
    \right)
    \nonumber\\
    &=
    -
    \frac{S\epsilon_0}{2}
    \sum_n
    \omega_{n,\chi}^{2}
    \left(
        \alpha_{n,\chi}\bm e_\chi
        -
        \alpha_{n,\chi}^{*}\bm e_\chi^{*}
    \right)^2
    \nonumber\\
    &=
    S\epsilon_0
    \sum_n
    \omega_{n,\chi}^{2}
    \alpha_{n,\chi}^{*}
    \alpha_{n,\chi}.
    \label{eq:Pi_energy}
\end{align}
Here we used Eq.~\eqref{eq:f_orthonormality} and the circular-vector
identities
\begin{equation}
    \bm e_\chi\cdot\bm e_{\chi'}
    =
    \delta_{\chi,-\chi'},
    \qquad
    \bm e_{-\chi}
    =
    \bm e_\chi^{*}.
    \label{eq:circular_unitvectors}
\end{equation}

The magnetic contribution is similarly
\begin{align}
    &
    \frac{S}{2\mu_0}
    \int dz\,
    \left(
        \frac{\partial\bm A_\chi}{\partial z}
    \right)^2
    \nonumber\\
    &=
    \frac{S\epsilon_0c^2}{2}
    \sum_{n,m}
    \int dz\,
    \frac{d}{dz}
    \left[
        \frac{f_{n,\chi}(z)}
        {\sqrt{\epsilon_\chi(z)}}
    \right]
    \frac{d}{dz}
    \left[
        \frac{f_{m,\chi}(z)}
        {\sqrt{\epsilon_\chi(z)}}
    \right]
    \nonumber\\
    &\qquad\times
    \left(
        \alpha_{n,\chi}\bm e_\chi
        +
        \alpha_{n,\chi}^{*}\bm e_\chi^{*}
    \right)
    \cdot
    \left(
        \alpha_{m,\chi}\bm e_\chi
        +
        \alpha_{m,\chi}^{*}\bm e_\chi^{*}
    \right)
    \nonumber\\
    &=
    -
    \frac{S\epsilon_0c^2}{2}
    \sum_{n,m}
    \int dz\,
    \frac{f_{n,\chi}(z)}
    {\sqrt{\epsilon_\chi(z)}}
    \frac{d^2}{dz^2}
    \left[
        \frac{f_{m,\chi}(z)}
        {\sqrt{\epsilon_\chi(z)}}
    \right]
    \nonumber\\
    &\qquad\times
    \left(
        \alpha_{n,\chi}\bm e_\chi
        +
        \alpha_{n,\chi}^{*}\bm e_\chi^{*}
    \right)
    \cdot
    \left(
        \alpha_{m,\chi}\bm e_\chi
        +
        \alpha_{m,\chi}^{*}\bm e_\chi^{*}
    \right)
    \nonumber\\
    &=
    \frac{S\epsilon_0}{2}
    \sum_{n,m}
    \omega_{m,\chi}^{2}
    \int dz\,
    f_{n,\chi}(z)
    f_{m,\chi}(z)
    \nonumber\\
    &\qquad\times
    \left(
        \alpha_{n,\chi}\bm e_\chi
        +
        \alpha_{n,\chi}^{*}\bm e_\chi^{*}
    \right)
    \cdot
    \left(
        \alpha_{m,\chi}\bm e_\chi
        +
        \alpha_{m,\chi}^{*}\bm e_\chi^{*}
    \right)
    \nonumber\\
    &=
    S\epsilon_0
    \sum_n
    \omega_{n,\chi}^{2}
    \alpha_{n,\chi}^{*}
    \alpha_{n,\chi}.
    \label{eq:A_energy}
\end{align}
In the second equality we integrated by parts and discarded the surface term
under the cavity boundary conditions. The eigenvalue equation
\eqref{eq:hermitian_wave_equation} and orthonormality
\eqref{eq:f_orthonormality} were then used.

The classical cavity Hamiltonian consequently becomes
\begin{equation}
    H_{\mathrm{cav}}
    =
    2S\epsilon_0
    \sum_{\chi,n}
    \omega_{n,\chi}^{2}
    \alpha_{n,\chi}^{*}
    \alpha_{n,\chi}.
    \label{eq:Hcav_alpha}
\end{equation}

Canonical quantization is implemented through
\begin{equation}
    \alpha_{n,\chi}
    =
    \sqrt{
        \frac{\hbar}
        {2\epsilon_0\omega_{n,\chi}S}
    }
    a_{n,\chi},
    \qquad
    \alpha_{n,\chi}^{*}
    \rightarrow
    \sqrt{
        \frac{\hbar}
        {2\epsilon_0\omega_{n,\chi}S}
    }
    a_{n,\chi}^{\dagger},
    \label{eq:alpha_quantization}
\end{equation}
with
\begin{equation}
    \left[
        a_{n,\chi},
        a_{n',\chi'}^\dagger
    \right]
    =
    \delta_{nn'}
    \delta_{\chi\chi'}.
\end{equation}
Up to the usual additive zero-point energy, the quantized cavity Hamiltonian
is therefore
\begin{equation}
    H_{\mathrm{cav}}
    =
    \sum_{\chi,n}
    \hbar\omega_{n,\chi}
    a_{n,\chi}^\dagger
    a_{n,\chi}.
    \label{eq:Hcav_quantized}
\end{equation}

The corresponding quantized vector potential is
\begin{align}
    \bm A_{\mathrm{cav}}(z)
    =
    \sum_{\chi,n}
    \sqrt{
        \frac{\hbar}
        {2\epsilon_0\epsilon_\chi(z)\omega_{n,\chi}S}
    }
    f_{n,\chi}(z)
    \left(
        a_{n,\chi}\bm e_\chi
        +
        a_{n,\chi}^\dagger\bm e_\chi^{*}
    \right),
    \label{eq:A_quantized}
\end{align}
and the conjugate momentum is
\begin{align}
    \bm\Pi_{\mathrm{cav}}(z)
    =
    i
    \sum_{\chi,n}
    \sqrt{
        \frac{
            \hbar\omega_{n,\chi}
            \epsilon_\chi(z)\epsilon_0
        }{
            2S
        }
    }
    f_{n,\chi}(z)
    \left(
        a_{n,\chi}^\dagger\bm e_\chi^{*}
        -
        a_{n,\chi}\bm e_\chi
    \right).
    \label{eq:Pi_quantized}
\end{align}


\section*{Supplementary Note 2: Minimal coupling to the cyclotron resonance}

We now derive the light--matter Hamiltonian describing the coupling between
the chiral cavity and a 2DEG subjected to a perpendicular external magnetic
field. We begin with $\bm B_{\mathrm{ext}}=+B\hat{\bm z}$ and introduce the
external vector potential $\bm A_0$ through
\begin{equation}
    \bm B_{\mathrm{ext}}
    =
    \bm\nabla\times\bm A_0.
\end{equation}
The total vector potential entering minimal coupling is
\begin{equation}
    \bm A
    =
    \bm A_0+\bm A_{\mathrm{cav}}.
\end{equation}
Because the external magnetic field is treated classically, it is convenient
to define the electronic mechanical momentum
\begin{equation}
    \bm\pi_j
    =
    \bm p_j+e\bm A_0(\bm r_j).
    \label{eq:mechanical_momentum}
\end{equation}

The electrons are confined to the $x$--$y$ plane and the 2DEG is located at
$z=z_{\mathrm{2DEG}}$. In Coulomb gauge and within the long-wavelength
approximation, the part of the Hamiltonian relevant to the spatially uniform
cyclotron response is
\begin{align}
    H
    ={}&
    H_{\mathrm{cav}}
    +
    \sum_{j=1}^{N}
    \frac{\bm\pi_j^2}{2m^\ast}
    \nonumber\\
    &+
    \frac{e}{m^\ast}
    \sum_{j=1}^{N}
    \bm\pi_j
    \cdot
    \bm A_{\mathrm{cav}}(z_{\mathrm{2DEG}})
    \nonumber\\
    &+
    \frac{e^2N}{2m^\ast}
    \bm A_{\mathrm{cav}}^{\,2}
    (z_{\mathrm{2DEG}}).
    \label{eq:Hamiltonian}
\end{align}
Here $m^\ast$ is the electron effective mass and $N$ is the number of
electrons within the quantization area $S$, with sheet density $n=N/S$.

Electron--electron interactions need not be neglected in the underlying
2DEG. For a parabolic dispersion and a spatially uniform electromagnetic
field, Kohn's theorem ensures that the center-of-mass cyclotron mode
decouples from the internal electronic degrees of freedom and retains the
cyclotron frequency. The derivation below therefore focuses directly on this
collective sector.

For $\bm B_{\mathrm{ext}}=+B\hat{\bm z}$, we take the positive cyclotron
frequency
\begin{equation}
    \omega_c
    =
    \frac{eB}{m^\ast}>0
\end{equation}
and define
\begin{align}
    c_j^\dagger
    &=
    -i
    \sqrt{
        \frac{1}
        {m^\ast\hbar\omega_c}
    }
    \pi_{j,-},
    \\
    c_j
    &=
    i
    \sqrt{
        \frac{1}
        {m^\ast\hbar\omega_c}
    }
    \pi_{j,+},
    \label{eq:c_operators}
\end{align}
where
\begin{equation}
    \pi_{j,\pm}
    =
    \bm e_{\mp}\cdot\bm\pi_j .
\end{equation}
Since
\begin{equation}
    [\pi_{j,+},\pi_{j,-}]
    =
    m^\ast\hbar\omega_c,
\end{equation}
the operators satisfy
\begin{equation}
    [c_j,c_j^\dagger]=1.
\end{equation}

We similarly introduce the circular projections of the cavity vector
potential,
\begin{equation}
    A_\pm
    =
    \bm e_{\mp}
    \cdot
    \bm A_{\mathrm{cav}}.
    \label{eq:A_projection_definition}
\end{equation}
Using
\begin{equation}
    \bm A_{\mathrm{cav}}^2
    =
    2A_+A_-,
\end{equation}
Eq.~\eqref{eq:Hamiltonian} becomes, up to an additive constant,
\begin{align}
    H
    ={}&
    H_{\mathrm{cav}}
    +
    \sum_{j=1}^{N}
    \hbar\omega_c
    c_j^\dagger c_j
    \nonumber\\
    &+
    \frac{e}{m^\ast}
    \sum_{j=1}^{N}
    \left[
        \pi_{j,-}
        A_+(z_{\mathrm{2DEG}})
        +
        A_-(z_{\mathrm{2DEG}})
        \pi_{j,+}
    \right]
    \nonumber\\
    &+
    \frac{e^2N}{m^\ast}
    A_+(z_{\mathrm{2DEG}})
    A_-(z_{\mathrm{2DEG}})
    \nonumber\\
    ={}&
    H_{\mathrm{cav}}
    +
    \sum_{j=1}^{N}
    \hbar\omega_c
    c_j^\dagger c_j
    \nonumber\\
    &+
    i
    \sqrt{
        \frac{
            e^2\hbar\omega_c
        }{
            m^\ast
        }
    }
    \sum_{j=1}^{N}
    \left[
        c_j^\dagger
        A_+(z_{\mathrm{2DEG}})
        -
        A_-(z_{\mathrm{2DEG}})
        c_j
    \right]
    \nonumber\\
    &+
    \frac{e^2N}{m^\ast}
    A_+(z_{\mathrm{2DEG}})
    A_-(z_{\mathrm{2DEG}}).
    \label{eq:Hplusminus}
\end{align}

Within the long-wavelength and linear-response regime, the cavity couples to
the collective center-of-mass cyclotron-resonance (CR) mode
\begin{equation}
    b
    =
    \frac{1}{\sqrt N}
    \sum_{j=1}^{N}
    c_j ,
    \label{eq:CR}
\end{equation}
which satisfies
\begin{equation}
    [b,b^\dagger]=1.
\end{equation}

From Eq.~\eqref{eq:A_quantized}, the two circular projections are
\begin{align}
    A_\pm(z)
    =
    \sum_n
    \Bigg[
        &
        \sqrt{
            \frac{\hbar}
            {2\epsilon_0
             \epsilon_\pm(z)
             \omega_{n,\pm}S}
        }
        f_{n,\pm}(z)
        a_{n,\pm}
        \nonumber\\
        &+
        \sqrt{
            \frac{\hbar}
            {2\epsilon_0
             \epsilon_\mp(z)
             \omega_{n,\mp}S}
        }
        f_{n,\mp}(z)
        a_{n,\mp}^\dagger
    \Bigg].
    \label{eq:Aplusminus}
\end{align}

Equation~\eqref{eq:Aplusminus} illustrates an important distinction. The
projection $A_\pm$ of the Hermitian electromagnetic field is not itself an
independent annihilation operator. It contains the annihilation operator of
one circular-polarization sector together with the creation operator associated with the
opposite circular component required by Hermiticity.

Substituting Eqs.~\eqref{eq:Aplusminus} and \eqref{eq:CR} into
Eq.~\eqref{eq:Hplusminus} gives
\begin{align}
    H
    ={}&
    \sum_n
    \hbar\omega_{n,+}
    a_{n,+}^\dagger a_{n,+}
    +
    \sum_n
    \hbar\omega_{n,-}
    a_{n,-}^\dagger a_{n,-}
    +
    \hbar\omega_c
    b^\dagger b
    \nonumber\\
    &+
    i\hbar
    \sum_n
    \bar g_{n,+}
    \left(
        a_{n,+}b^\dagger
        -
        a_{n,+}^\dagger b
    \right)
    \nonumber\\
    &+
    i\hbar
    \sum_n
    \bar g_{n,-}
    \left(
        a_{n,-}^\dagger b^\dagger
        -
        a_{n,-}b
    \right)
    \nonumber\\
    &+
    \hbar
    \sum_{n,n'}
    \frac{
        \bar g_{n,+}\bar g_{n',-}
    }{
        \omega_c
    }
    \left(
        a_{n,+}a_{n',-}
        +
        a_{n,+}^\dagger
        a_{n',-}^\dagger
    \right)
    \nonumber\\
    &+
    \hbar
    \sum_{n,n'}
    \frac{
        \bar g_{n,+}\bar g_{n',+}
    }{
        \omega_c
    }
    a_{n,+}^\dagger a_{n',+}
    \nonumber\\
    &+
    \hbar
    \sum_{n,n'}
    \frac{
        \bar g_{n,-}\bar g_{n',-}
    }{
        \omega_c
    }
    a_{n,-}^\dagger a_{n',-},
    \label{eq:chiral_H}
\end{align}
up to an irrelevant additive constant.

The field-dependent collective coupling strengths are
\begin{equation}
    \bar g_{n,\pm}
    =
    \sqrt{
        \frac{
            ne^2\omega_c
        }{
            2m^\ast
            \epsilon_0
            \epsilon_\pm(z_{\mathrm{2DEG}})
            \omega_{n,\pm}
        }
    }
    f_{n,\pm}(z_{\mathrm{2DEG}}).
    \label{eq:g_general}
\end{equation}

Equation~\eqref{eq:chiral_H} reveals the physical separation produced by the
circular basis. For $\bm B_{\mathrm{ext}}=+B\hat{\bm z}$, the
$\chi=+1$ cavity sector couples to the CR through excitation-exchange terms,
whereas the $\chi=-1$ sector enters through anomalous pair-creation and
pair-annihilation terms. The diamagnetic interaction also contains
number-conserving shifts within each circular-polarization sector and anomalous coupling
between opposite circular polarizations.

This is the same circular decomposition underlying the cyclotron-active and
cyclotron-inactive sectors of conventional Landau-polariton cavities and the
vacuum Bloch--Siegert response~\cite{Li2018}.


\subsection*{Two-polarization reference Hamiltonian}

For the comparison used in the main text, we retain one degenerate cavity
mode of each circular polarization and denote their annihilation operators by $a_+$ and
$a_-$. For $\bm B_{\mathrm{ext}}=+B\hat{\bm z}$,
Eq.~\eqref{eq:chiral_H} reduces to
\begin{align}
    H_{\mathrm{achiral}}^{(+B)}
    ={}&
    \hbar\widetilde\omega_0
    \left(
        a_+^\dagger a_+
        +
        a_-^\dagger a_-
    \right)
    +
    \hbar\omega_c b^\dagger b
    \nonumber\\
    &+
    i\hbar\bar g
    \left(
        a_+b^\dagger
        -
        a_+^\dagger b
    \right)
    \nonumber\\
    &+
    i\hbar\bar g
    \left(
        a_-^\dagger b^\dagger
        -
        a_-b
    \right)
    \nonumber\\
    &+
    \hbar D
    \left(
        a_+a_-
        +
        a_+^\dagger a_-^\dagger
    \right),
    \label{eq:H_twohelicity_plus}
\end{align}
where the common number-conserving diamagnetic contribution has been absorbed
into
\begin{equation}
    \widetilde\omega_0
    =
    \omega_0+D,
    \qquad
    D
    =
    \frac{\bar g^2}{\omega_c}.
    \label{eq:D_definition}
\end{equation}

Reversing the magnetic field reverses the cyclotron motion while leaving the
two cavity circular polarizations unchanged. Their roles are therefore exchanged:
\begin{align}
    H_{\mathrm{achiral}}^{(-B)}
    ={}&
    \hbar\widetilde\omega_0
    \left(
        a_+^\dagger a_+
        +
        a_-^\dagger a_-
    \right)
    +
    \hbar\omega_c b^\dagger b
    \nonumber\\
    &+
    i\hbar\bar g
    \left(
        a_-b^\dagger
        -
        a_-^\dagger b
    \right)
    \nonumber\\
    &+
    i\hbar\bar g
    \left(
        a_+^\dagger b^\dagger
        -
        a_+b
    \right)
    \nonumber\\
    &+
    \hbar D
    \left(
        a_+a_-
        +
        a_+^\dagger a_-^\dagger
    \right).
    \label{eq:H_twohelicity_minus}
\end{align}

Thus, in a non-chiral cavity supporting both circular polarizations, reversal of $B$
does not eliminate the resonant interaction. It simply exchanges which
circular cavity mode participates in resonant excitation exchange and which
one participates in the anomalous interaction. This is the two-polarization
reference used for Figs.~2b and 3c,d of the main text.


\subsection*{Single-polarization limit}

We now take the ideal chiral limit relevant to the main result. We assume
that only one cavity mode lies in the frequency window of interest and that
this mode is RCP polarized,
\begin{equation}
    \chi=-1.
\end{equation}
We denote its frequency by $\omega_0$ and its annihilation operator by
$a_-$. The vector potential becomes
\begin{equation}
    \bm A_{\mathrm{cav}}(z)
    =
    \sqrt{
        \frac{\hbar}
        {2\epsilon_0
         \epsilon_{\mathrm{bg}}
         \omega_0S}
    }
    f(z)
    \left(
        a_-\bm e_-
        +
        a_-^\dagger\bm e_-^{*}
    \right).
    \label{eq:chiral_A}
\end{equation}

For $\bm B_{\mathrm{ext}}=+B\hat{\bm z}$, the supported RCP mode belongs to
the anomalous interaction sector. Equation~\eqref{eq:chiral_H} therefore
reduces to
\begin{align}
    H
    ={}&
    \hbar\omega_0
    a_-^\dagger a_-
    +
    \hbar\omega_c
    b^\dagger b
    \nonumber\\
    &+
    i\hbar\bar g
    \left(
        a_-^\dagger b^\dagger
        -
        a_-b
    \right)
    +
    \hbar
    \frac{\bar g^2}{\omega_c}
    a_-^\dagger a_-
    \nonumber\\
    ={}&
    \hbar\widetilde\omega_0
    a_-^\dagger a_-
    +
    \hbar\omega_c
    b^\dagger b
    +
    i\hbar\bar g
    \left(
        a_-^\dagger b^\dagger
        -
        a_-b
    \right),
    \label{eq:H_counter}
\end{align}
with
\begin{equation}
    \boxed{
    \widetilde\omega_0
    =
    \omega_0
    +
    \frac{\bar g^2}{\omega_c}.
    }
    \label{eq:omega_tilde}
\end{equation}

A key consequence of the single-polarization limit is visible directly from the
diamagnetic term. Since
\begin{equation}
    \bm e_-\cdot\bm e_-=0,
\end{equation}
one finds
\begin{equation}
    \bm A_{\mathrm{cav}}^2
    \propto
    2a_-^\dagger a_-+1,
\end{equation}
so the diamagnetic interaction contains no $a_-^2$ or
$a_-^{\dagger2}$ terms. It therefore shifts the cavity frequency but does not
generate independent photonic squeezing.

Equation~\eqref{eq:H_counter} is a two-mode squeezing Hamiltonian. It does
not hybridize the one-excitation cavity and CR states through resonant
exchange and therefore produces no conventional polaritonic avoided
crossing. Instead, the interaction connects the bare vacuum to states
containing correlated cavity and CR excitations, producing quantum-vacuum
dressing and the vacuum Bloch--Siegert spectral shift.


\subsection*{Magnetic-field reversal}

We finally reverse the external magnetic field,
\begin{equation}
    \bm B_{\mathrm{ext}}
    =
    -B\hat{\bm z},
\end{equation}
while keeping
\begin{equation}
    \omega_c
    =
    \frac{e|B|}{m^\ast}>0 .
\end{equation}
The sign of the field is therefore carried by the sense of the cyclotron
motion rather than by allowing $\omega_c$ itself to become negative.

For the reversed field,
\begin{equation}
    [\pi_{j,+},\pi_{j,-}]
    =
    -m^\ast\hbar\omega_c .
\end{equation}
To preserve
\begin{equation}
    [c_j,c_j^\dagger]=1,
\end{equation}
the circular mechanical momenta appearing in the ladder operators are
exchanged:
\begin{align}
    c_j^\dagger
    &=
    -i
    \sqrt{
        \frac{1}
        {m^\ast\hbar\omega_c}
    }
    \pi_{j,+},
    \\
    c_j
    &=
    i
    \sqrt{
        \frac{1}
        {m^\ast\hbar\omega_c}
    }
    \pi_{j,-}.
    \label{eq:c_operators_negativeB}
\end{align}

For the same fixed RCP cavity mode, the interaction is consequently converted
into resonant excitation exchange,
\begin{equation}
    H_{\bm B_{\mathrm{ext}}=-B\hat{\bm z}}
    =
    \hbar\widetilde\omega_0
    a_-^\dagger a_-
    +
    \hbar\omega_c
    b^\dagger b
    +
    i\hbar\bar g
    \left(
        a_-b^\dagger
        -
        a_-^\dagger b
    \right).
    \label{eq:H_co}
\end{equation}

This Hamiltonian has the structure of two linearly coupled bosonic
oscillators. Importantly, its number-conserving form has not been obtained by
applying the rotating-wave approximation. It follows directly from the
circular polarization of the cavity mode together with the direction of the cyclotron
motion.

Defining
\begin{equation}
    N_{\mathrm{exc}}
    =
    a_-^\dagger a_-
    +
    b^\dagger b,
\end{equation}
one has
\begin{equation}
    \left[
        H_{\bm B_{\mathrm{ext}}=-B\hat{\bm z}},
        N_{\mathrm{exc}}
    \right]
    =
    0 .
\end{equation}
The zero-excitation state is therefore an exact eigenstate. Moreover, because
the minimal-coupling diamagnetic contribution keeps the excitation spectrum
stable, this state is the exact ground state of the single-polarization model.
Consequently,
\begin{equation}
    \langle a_-^\dagger a_-\rangle
    =
    \langle b^\dagger b\rangle
    =
    \langle a_-b\rangle
    =
    0
\end{equation}
even though the excitation spectrum exhibits a large Rabi splitting.

The two magnetic-field directions can therefore be summarized compactly as
\begin{equation}
    H_{\mathrm{chiral}}
    =
    \hbar\widetilde\omega_0
    a_-^\dagger a_-
    +
    \hbar\omega_c
    b^\dagger b
    +
    \begin{cases}
        i\hbar\bar g
        \left(
            a_-b^\dagger
            -
            a_-^\dagger b
        \right),
        & B<0,
        \\[2mm]
        i\hbar\bar g
        \left(
            a_-^\dagger b^\dagger
            -
            a_-b
        \right),
        & B>0.
    \end{cases}
    \label{eq:total_H}
\end{equation}
Here $B$ denotes the signed magnetic field used throughout the main text,
while $\omega_c=e|B|/m^\ast$ remains positive.

Equation~\eqref{eq:total_H} makes the central distinction explicit. For
$B<0$, the complete single-polarization Hamiltonian is number conserving and
produces polariton formation while leaving the bare vacuum unchanged. For
$B>0$, the same fixed cavity mode couples through anomalous pair processes,
so the conventional resonant anticrossing disappears while the ground state
becomes quantum mechanically dressed.


\section*{Supplementary Note 3: Diagonalization and ground-state properties}

We now diagonalize the single-polarization and two-polarization Hamiltonians derived
in Supplementary Note~2 and obtain the ground-state quantities used in the
main text. We keep
\begin{equation}
    \omega_c=\frac{e|B|}{m^\ast}>0,
\end{equation}
throughout, while the sign of $B$ determines the sense of the cyclotron
rotation relative to the fixed circular polarization of the cavity mode.

The diagonalization of quadratic bosonic Hamiltonians by ordinary mode
rotations or bosonic Bogoliubov transformations is standard
~\cite{Hopfield1958,CiutiPRB2005}. We therefore state below the corresponding
transformations and results for the Hamiltonians relevant to the present
system.


\subsection*{Single-polarization cavity for \texorpdfstring{$B<0$}{B<0}}

For $B<0$, the fixed RCP cavity mode rotates in the same sense as the
cyclotron motion. The Hamiltonian is
\begin{equation}
    H(-B)
    =
    \hbar\widetilde\omega_0
    a_-^\dagger a_-
    +
    \hbar\omega_c
    b^\dagger b
    +
    i\hbar\bar g
    \left(
        a_-b^\dagger-a_-^\dagger b
    \right).
    \label{eq:H_minusB_diag}
\end{equation}
Here
\begin{equation}
    \widetilde\omega_0
    =
    \omega_0+D,
    \qquad
    D=\frac{\bar g^2}{\omega_c}.
\end{equation}

After the phase transformation
\begin{equation}
    \widetilde b=-ib,
\end{equation}
Eq.~\eqref{eq:H_minusB_diag} becomes the standard Hamiltonian of two linearly
coupled bosonic modes,
\begin{equation}
    H(-B)
    =
    \hbar\widetilde\omega_0
    a_-^\dagger a_-
    +
    \hbar\omega_c
    \widetilde b^\dagger\widetilde b
    +
    \hbar\bar g
    \left(
        a_-^\dagger\widetilde b
        +
        \widetilde b^\dagger a_-
    \right).
    \label{eq:H_minusB_real}
\end{equation}

It is diagonalized by the rotation
\begin{align}
    P_{\mathrm{UP}}
    &=
    \cos\theta\,a_-
    +
    \sin\theta\,\widetilde b,
    \\
    P_{\mathrm{LP}}
    &=
    -\sin\theta\,a_-
    +
    \cos\theta\,\widetilde b,
    \label{eq:polariton_rotation}
\end{align}
where
\begin{equation}
    \tan(2\theta)
    =
    \frac{2\bar g}
    {\widetilde\omega_0-\omega_c}.
    \label{eq:mixing_angle}
\end{equation}
The Hamiltonian becomes
\begin{equation}
    H(-B)
    =
    \hbar\Omega_{\mathrm{UP}}
    P_{\mathrm{UP}}^\dagger P_{\mathrm{UP}}
    +
    \hbar\Omega_{\mathrm{LP}}
    P_{\mathrm{LP}}^\dagger P_{\mathrm{LP}},
    \label{eq:H_minusB_diagonal}
\end{equation}
with
\begin{equation}
    \Omega_{\mathrm{UP/LP}}
    =
    \frac{
        \widetilde\omega_0+\omega_c
    }{2}
    \pm
    \sqrt{
        \left(
            \frac{
                \widetilde\omega_0-\omega_c
            }{2}
        \right)^2
        +
        \bar g^2
    }.
    \label{eq:UPLP_frequencies}
\end{equation}

The lower mode remains positive because
\begin{align}
    \widetilde\omega_0\omega_c-\bar g^2
    &=
    \left(
        \omega_0+\frac{\bar g^2}{\omega_c}
    \right)
    \omega_c-\bar g^2
    \nonumber\\
    &=
    \omega_0\omega_c
    >
    0.
    \label{eq:stability_minusB}
\end{align}

More importantly for the ground state,
\begin{equation}
    \left[
        H(-B),
        a_-^\dagger a_-+b^\dagger b
    \right]
    =
    0.
    \label{eq:number_conservation_minusB}
\end{equation}
Since the transformation in Eq.~\eqref{eq:polariton_rotation} only mixes
annihilation operators, the polariton vacuum is identical to the bare
cavity--CR vacuum. The exact ground state is therefore
\begin{equation}
    \boxed{
    |G(-B)\rangle
    =
    |0_{a_-},0_b\rangle .
    }
    \label{eq:G_minusB}
\end{equation}
It follows immediately that
\begin{equation}
    \boxed{
    \langle a_-^\dagger a_-\rangle_{-B}
    =
    \langle b^\dagger b\rangle_{-B}
    =
    \langle a_-b\rangle_{-B}
    =
    0.
    }
    \label{eq:GS_minusB}
\end{equation}

Thus the upper and lower polariton branches arise from a purely
number-conserving hybridization and coexist with an unchanged bare ground
state.


\subsection*{Single-polarization cavity for \texorpdfstring{$B>0$}{B>0}}

For $B>0$, the cyclotron motion reverses while the RCP cavity polarization remains
fixed. The Hamiltonian becomes
\begin{equation}
    H(+B)
    =
    \hbar\widetilde\omega_0
    a_-^\dagger a_-
    +
    \hbar\omega_c
    b^\dagger b
    +
    i\hbar\bar g
    \left(
        a_-^\dagger b^\dagger-a_-b
    \right).
    \label{eq:H_plusB_diag}
\end{equation}

Using again
\begin{equation}
    \widetilde b=-ib,
\end{equation}
we obtain
\begin{equation}
    H(+B)
    =
    \hbar\widetilde\omega_0
    a_-^\dagger a_-
    +
    \hbar\omega_c
    \widetilde b^\dagger\widetilde b
    +
    \hbar\bar g
    \left(
        a_-^\dagger\widetilde b^\dagger
        +
        a_-\widetilde b
    \right).
    \label{eq:H_plusB_real}
\end{equation}

This is the standard two-mode bosonic squeezing Hamiltonian. Its
Bogoliubov diagonalization is obtained with
\begin{align}
    P_1
    &=
    \cosh r_B\,a_-
    +
    \sinh r_B\,\widetilde b^\dagger,
    \\
    P_2
    &=
    \cosh r_B\,\widetilde b
    +
    \sinh r_B\,a_-^\dagger,
    \label{eq:Bogoliubov_single}
\end{align}
where~\cite{CiutiPRB2005}
\begin{equation}
    \tanh(2r_B)
    =
    \frac{
        2\bar g
    }{
        \widetilde\omega_0+\omega_c
    }.
    \label{eq:rB_definition}
\end{equation}

Defining
\begin{equation}
    \Delta_B
    =
    \sqrt{
        \left(
            \widetilde\omega_0+\omega_c
        \right)^2
        -
        4\bar g^2
    },
    \label{eq:DeltaB}
\end{equation}
the diagonal Hamiltonian is
\begin{equation}
    H(+B)
    =
    \hbar\Omega_1
    P_1^\dagger P_1
    +
    \hbar\Omega_2
    P_2^\dagger P_2
    +
    E_G(+B),
    \label{eq:H_plusB_diagonal}
\end{equation}
with
\begin{align}
    \Omega_1
    &=
    \frac{1}{2}
    \left[
        \Delta_B
        +
        \widetilde\omega_0-\omega_c
    \right],
    \\
    \Omega_2
    &=
    \frac{1}{2}
    \left[
        \Delta_B
        -
        \widetilde\omega_0+\omega_c
    \right].
    \label{eq:plusB_frequencies}
\end{align}

The stability condition is again guaranteed by the diamagnetic contribution,
since
\begin{equation}
    \widetilde\omega_0\omega_c-\bar g^2
    =
    \omega_0\omega_c
    >
    0.
    \label{eq:stability_plusB}
\end{equation}

The ground state is defined by
\begin{equation}
    P_1|G(+B)\rangle
    =
    P_2|G(+B)\rangle
    =
    0.
\end{equation}
The corresponding standard two-mode squeezed vacuum is
~\cite{CiutiPRB2005}
\begin{equation}
    \boxed{
    |G(+B)\rangle
    =
    \frac{1}{\cosh r_B}
    \sum_{n=0}^{\infty}
    \left(
        -i\tanh r_B
    \right)^n
    |n_{a_-},n_b\rangle .
    }
    \label{eq:G_plusB}
\end{equation}

The ground-state observables required in the main text follow directly from
the standard two-mode squeezed-state relations:
\begin{equation}
    \langle a_-^\dagger a_-\rangle_{+B}
    =
    \sinh^2r_B,
    \label{eq:nph_plusB}
\end{equation}
\begin{equation}
    \langle b^\dagger b\rangle_{+B}
    =
    \sinh^2r_B,
    \label{eq:nCR_plusB}
\end{equation}
and
\begin{equation}
    \left|
        \langle a_-b\rangle_{+B}
    \right|
    =
    \frac{1}{2}\sinh(2r_B).
    \label{eq:corr_plusB}
\end{equation}
Equivalently,
\begin{equation}
    \langle a_-^\dagger a_-\rangle_{+B}
    =
    \frac{1}{2}
    \left[
        \frac{
            \widetilde\omega_0+\omega_c
        }{
            \Delta_B
        }
        -1
    \right],
    \label{eq:nph_plusB_closed}
\end{equation}
\begin{equation}
    \langle b^\dagger b\rangle_{+B}
    =
    \frac{1}{2}
    \left[
        \frac{
            \widetilde\omega_0+\omega_c
        }{
            \Delta_B
        }
        -1
    \right],
    \label{eq:nCR_plusB_closed}
\end{equation}
and
\begin{equation}
    \left|
        \langle a_-b\rangle_{+B}
    \right|
    =
    \frac{\bar g}{\Delta_B}.
    \label{eq:corr_plusB_closed}
\end{equation}

Unlike the $B<0$ case, the Bogoliubov transformation mixes creation and
annihilation operators. The bare vacuum is therefore not the ground state,
and the interaction generates finite virtual populations and anomalous
cavity--CR correlations.


\subsection*{Two-polarization reference cavity}

For comparison with the single-polarization cavity, we consider a non-chiral
cavity supporting two degenerate circular modes,
\begin{equation}
    a_+,
    \qquad
    a_-,
\end{equation}
with
\begin{equation}
    \chi=+1\equiv\mathrm{LCP},
    \qquad
    \chi=-1\equiv\mathrm{RCP}.
\end{equation}
The two modes have the same bare cavity frequency $\omega_0$ and the same
light--matter coupling strength $\bar g$.

For $B>0$, the LCP mode participates in the co-rotating excitation-exchange
channel with the cyclotron resonance, whereas the RCP mode enters through
the counter-rotating channel. The Hamiltonian is
\begin{align}
    H_{\mathrm{achiral}}(+B)
    ={}&
    \hbar\widetilde\omega_0
    \left(
        a_+^\dagger a_+
        +
        a_-^\dagger a_-
    \right)
    +
    \hbar\omega_c b^\dagger b
    \nonumber\\
    &+
    i\hbar\bar g
    \left(
        a_+b^\dagger
        -
        a_+^\dagger b
    \right)
    \nonumber\\
    &+
    i\hbar\bar g
    \left(
        a_-^\dagger b^\dagger
        -
        a_-b
    \right)
    \nonumber\\
    &+
    \hbar D
    \left(
        a_+a_-
        +
        a_+^\dagger a_-^\dagger
    \right),
    \label{eq:H2h_plusB}
\end{align}
where the number-conserving contribution of the diamagnetic term has been
absorbed into
\begin{equation}
    \widetilde\omega_0=\omega_0+D.
\end{equation}

For $B<0$, the cyclotron motion reverses and the two circular cavity
polarizations exchange their roles,
\begin{align}
    H_{\mathrm{achiral}}(-B)
    ={}&
    \hbar\widetilde\omega_0
    \left(
        a_+^\dagger a_+
        +
        a_-^\dagger a_-
    \right)
    +
    \hbar\omega_c b^\dagger b
    \nonumber\\
    &+
    i\hbar\bar g
    \left(
        a_-b^\dagger
        -
        a_-^\dagger b
    \right)
    \nonumber\\
    &+
    i\hbar\bar g
    \left(
        a_+^\dagger b^\dagger
        -
        a_+b
    \right)
    \nonumber\\
    &+
    \hbar D
    \left(
        a_+a_-
        +
        a_+^\dagger a_-^\dagger
    \right).
    \label{eq:H2h_minusB}
\end{align}

Equations~\eqref{eq:H2h_plusB} and \eqref{eq:H2h_minusB} show that reversing
the magnetic field does not remove polariton formation in the two-polarization
reference cavity. Instead, it exchanges the circular polarization that
participates in the co-rotating light--matter interaction, as in the
conventional polarization-resolved Landau-polariton response~\cite{Li2018}.

Because the two magnetic-field directions are related by
\begin{equation}
    a_+\longleftrightarrow a_-,
\end{equation}
it is sufficient to present the Bogoliubov diagonalization explicitly for
$B>0$.

We collect the three bare annihilation operators into
\begin{equation}
    \bm a
    =
    \begin{pmatrix}
        a_+\\
        a_-\\
        b
    \end{pmatrix}.
    \label{eq:a_vector_2h}
\end{equation}
The Hamiltonian can then be written in the general quadratic bosonic form
\begin{equation}
    H_{\mathrm{achiral}}(+B)
    =
    \hbar\bm a^\dagger h\,\bm a
    +
    \frac{\hbar}{2}
    \left(
        \bm a^\dagger\Delta\,\bm a^{\dagger T}
        +
        \bm a^T\Delta^\ast\,\bm a
    \right),
    \label{eq:H_general_2h}
\end{equation}
up to an additive constant, with
\begin{equation}
    h
    =
    \begin{pmatrix}
        \widetilde\omega_0 & 0 & -i\bar g\\
        0 & \widetilde\omega_0 & 0\\
        i\bar g & 0 & \omega_c
    \end{pmatrix},
    \label{eq:h_2h}
\end{equation}
and
\begin{equation}
    \Delta
    =
    \begin{pmatrix}
        0 & D & 0\\
        D & 0 & i\bar g\\
        0 & i\bar g & 0
    \end{pmatrix}.
    \label{eq:Delta_2h}
\end{equation}

The Heisenberg equations for the doubled bare-mode vector are therefore
\begin{equation}
    i\frac{d}{dt}
    \begin{pmatrix}
        \bm a\\
        \bm a^\dagger
    \end{pmatrix}
    =
    \mathcal M
    \begin{pmatrix}
        \bm a\\
        \bm a^\dagger
    \end{pmatrix},
    \label{eq:Heisenberg_2h}
\end{equation}
where the bosonic dynamical matrix is
\begin{equation}
    \boxed{
    \mathcal M
    =
    \begin{pmatrix}
        h & \Delta\\
        -\Delta^\ast & -h^\ast
    \end{pmatrix}.
    }
    \label{eq:M_2h}
\end{equation}

The eigenvalues of $\mathcal M$ occur in positive--negative pairs,
\begin{equation}
    \{
        \Omega_1,\Omega_2,\Omega_3,
        -\Omega_1,-\Omega_2,-\Omega_3
    \},
\end{equation}
and the positive frequencies $\Omega_\nu$ define the three physical
normal modes.

We introduce the Bogoliubov operators
\begin{equation}
    P_\nu
    =
    \sum_{i=1}^{3}
    \left(
        w_{\nu i}a_i
        +
        y_{\nu i}a_i^\dagger
    \right),
    \qquad
    \nu=1,2,3,
    \label{eq:Bogoliubov_2h}
\end{equation}
where
\begin{equation}
    (a_1,a_2,a_3)
    =
    (a_+,a_-,b).
\end{equation}
The transformation preserves the bosonic commutation relations provided
\begin{align}
    W^\dagger W-Y^\dagger Y &= \mathbb I,
    \\
    WY^T-YW^T &=0,
    \label{eq:paraunitary_2h}
\end{align}
where
\begin{equation}
    W_{\nu i}=w_{\nu i},
    \qquad
    Y_{\nu i}=y_{\nu i}.
\end{equation}

For each positive-frequency mode, the corresponding eigenvector of
$\mathcal M$ may be written as
\begin{equation}
    \bm\phi_\nu
    =
    \begin{pmatrix}
        \bm w_\nu^\ast\\
        -\bm y_\nu^\ast
    \end{pmatrix},
    \label{eq:eigenvector_2h}
\end{equation}
with bosonic normalization
\begin{equation}
    \bm\phi_\nu^\dagger
    \Sigma
    \bm\phi_{\nu'}
    =
    \delta_{\nu\nu'},
    \qquad
    \Sigma=
    \begin{pmatrix}
        \mathbb I_3 & 0\\
        0 & -\mathbb I_3
    \end{pmatrix}.
    \label{eq:norm_2h}
\end{equation}
Equivalently,
\begin{equation}
    \sum_i
    \left(
        |w_{\nu i}|^2-|y_{\nu i}|^2
    \right)
    =1.
\end{equation}

The inverse Bogoliubov transformation is
\begin{equation}
    \boxed{
    a_i
    =
    \sum_{\nu=1}^{3}
    \left(
        w_{\nu i}^\ast P_\nu
        -
        y_{\nu i}P_\nu^\dagger
    \right).
    }
    \label{eq:inverse_Bogoliubov_2h}
\end{equation}
The diagonal Hamiltonian therefore takes the form
\begin{equation}
    H_{\mathrm{achiral}}(+B)
    =
    \sum_{\nu=1}^{3}
    \hbar\Omega_\nu
    P_\nu^\dagger P_\nu
    +
    E_G^{\mathrm{2h}}(+B).
    \label{eq:H2h_diagonal}
\end{equation}

The interacting ground state is defined by
\begin{equation}
    P_\nu
    |G_{\mathrm{2h}}(+B)\rangle
    =
    0,
    \qquad
    \nu=1,2,3.
    \label{eq:G2h_definition}
\end{equation}
The bare-mode virtual populations then follow directly from the anomalous
Bogoliubov coefficients,
\begin{equation}
    \boxed{
    \langle a_i^\dagger a_i\rangle
    =
    \sum_{\nu=1}^{3}
    |y_{\nu i}|^2 .
    }
    \label{eq:population_2h}
\end{equation}
In particular,
\begin{align}
    n_+
    &=
    \langle a_+^\dagger a_+\rangle,
    \\
    n_-
    &=
    \langle a_-^\dagger a_-\rangle,
    \\
    n_{\mathrm{CR}}
    &=
    \langle b^\dagger b\rangle,
\end{align}
and the total cavity population plotted in the main text is
\begin{equation}
    n_{\mathrm{ph}}
    =
    n_++n_-.
    \label{eq:nph_2h}
\end{equation}

The same inverse transformation gives the anomalous ground-state
correlations,
\begin{equation}
    \boxed{
    \langle a_i a_j\rangle
    =
    -
    \sum_{\nu=1}^{3}
    w_{\nu i}^\ast y_{\nu j}.
    }
    \label{eq:anomalous_2h}
\end{equation}
The normal correlations are
\begin{equation}
    \boxed{
    \langle a_i^\dagger a_j\rangle
    =
    \sum_{\nu=1}^{3}
    y_{\nu i}^\ast y_{\nu j}.
    }
    \label{eq:normal_2h}
\end{equation}

Hence the two circular cavity--CR anomalous correlations shown in the main
text are
\begin{equation}
    \left|
        \langle a_-b\rangle
    \right|
    \qquad\mathrm{and}\qquad
    \left|
        \langle a_+b\rangle
    \right|.
    \label{eq:2h_correlations}
\end{equation}

For $B<0$, the identical procedure applies after exchanging
$a_+\leftrightarrow a_-$. The positive-frequency spectrum is therefore
unchanged, while the circular polarization associated with the co-rotating and
counter-rotating interaction channels is reversed.

\section*{Supplementary Note 4: Transfer-matrix calculations}

The electromagnetic response of the cavity is calculated using the
transfer-matrix method (TMM) for stratified media~\cite{Yeh1988}. At normal
incidence, right- and left-circularly polarized fields propagate independently,
so the RCP and LCP components can be treated as separate optical channels. Throughout
this work we use
\begin{equation}
    \mathrm{RCP}:\chi=-1,
    \qquad
    \mathrm{LCP}:\chi=+1.
    \label{eq:TMM_helicity}
\end{equation}


\subsection*{Cavity and chiral mirrors}

The cavity used throughout the manuscript consists of the one-dimensional
stack
\begin{equation}
    M_\chi
    \;|\;
    \mathrm{Air}
    \;|\;
    \mathrm{QW}
    \;|\;
    \mathrm{Si}
    \;|\;
    \mathrm{Air}
    \;|\;
    M_\chi ,
    \label{eq:TMM_stack}
\end{equation}
where $M_\chi$ denotes a chiral mirror. The corresponding layer thicknesses
are
\begin{equation}
    d_{\mathrm M}=45~\mu\mathrm{m},
    \qquad
    d_{\mathrm{air}}=96~\mu\mathrm{m},
    \qquad
    d_{\mathrm{QW}}=2.4~\mu\mathrm{m},
    \qquad
    d_{\mathrm{Si}}=150~\mu\mathrm{m},
    \label{eq:TMM_dimensions}
\end{equation}
and we take
\begin{equation}
    n_{\mathrm{air}}=1,
    \qquad
    n_{\mathrm{Si}}=3.4,
    \qquad
    \epsilon_{\mathrm{bg}}=3.6^2=12.96.
    \label{eq:TMM_indices}
\end{equation}
Before the free-carrier response of the 2DEG is introduced, the quantum-well
membrane is treated as a passive dielectric with
\begin{equation}
    n_{\mathrm{QW}}^{(0)}
    =
    \sqrt{\epsilon_{\mathrm{bg}}}
    =
    3.6.
    \label{eq:TMM_passive_QW}
\end{equation}
The passive dielectric loading shifts the selected RCP cavity resonance
relative to the membrane-free structure. For the parameters above, the RCP
resonance occurs at approximately
\begin{equation}
    f_0
    \simeq
    0.593~\mathrm{THz},
    \label{eq:TMM_loaded_f0}
\end{equation}
with a quality factor $Q\simeq63$.

The effective chiral-mirror response is motivated by the InSb
photonic-crystal cavity demonstrated in Ref.~\cite{Kulkarni2026}.
For $B\approx0.3$ T, the CRI (RCP) response of InSb has negligible absorption
near the operating frequency and a real permittivity approaching that
of Si, whereas the CRA (LCP) response is strongly absorptive near the
cyclotron resonance. We represent this experimentally demonstrated
contrast by the idealized effective indices
\begin{equation}
    n_{\mathrm M}^{\mathrm{RCP}}=n_{\mathrm{Si}},
    \qquad
    n_{\mathrm M}^{\mathrm{LCP}}=n_{\mathrm{Si}}+5i.
\end{equation}
The value $5i$ is chosen to realize the strongly absorbing limit and
is not intended as a microscopic fit to InSb.
The imaginary part in the LCP channel is introduced only to represent the
strongly suppressed LCP response and is not intended as a microscopic fit
to the magnetoplasma mirror. The mirror response is kept fixed throughout the
magnetic-field sweep.

For a homogeneous layer $j$ of refractive index $n_j$, thickness $d_j$ and
phase accumulation
\begin{equation}
    \delta_j
    =
    \frac{\omega}{c}n_jd_j,
\end{equation}
the normal-incidence characteristic matrix is
\begin{equation}
    \mathcal M_j
    =
    \begin{pmatrix}
        \cos\delta_j
        &
        \displaystyle
        \frac{-i}{n_j}\sin\delta_j
        \\[2mm]
        -i n_j\sin\delta_j
        &
        \cos\delta_j
    \end{pmatrix}.
    \label{eq:TMM_layer_matrix}
\end{equation}

For each circular-polarization channel, the transfer matrix of the passively loaded cavity is
obtained by multiplying the matrices of the individual layers in their
physical order,
\begin{equation}
    \mathcal M_\chi
    =
    \mathcal M_{M_\chi}
    \mathcal M_{\mathrm{air}}
    \mathcal M_{\mathrm{QW}}^{(0)}
    \mathcal M_{\mathrm{Si}}
    \mathcal M_{\mathrm{air}}
    \mathcal M_{M_\chi}
    =
    \begin{pmatrix}
        M_{11} & M_{12}\\
        M_{21} & M_{22}
    \end{pmatrix}.
    \label{eq:TMM_total_matrix}
\end{equation}

For incident and output refractive indices $n_{\mathrm{in}}$ and
$n_{\mathrm{out}}$, respectively, the complex transmission amplitude is
\begin{equation}
    t_\chi
    =
    \frac{
        2n_{\mathrm{in}}
    }{
        n_{\mathrm{in}}M_{11}
        +
        n_{\mathrm{in}}n_{\mathrm{out}}M_{12}
        +
        M_{21}
        +
        n_{\mathrm{out}}M_{22}
    }.
    \label{eq:TMM_t}
\end{equation}
In the present structure both external media are air,
$n_{\mathrm{in}}=n_{\mathrm{out}}=1$, and therefore the power transmittance
reduces to
\begin{equation}
    T_\chi
    =
    |t_\chi|^2.
    \label{eq:TMM_T}
\end{equation}
For a linearly polarized incident field, the two circular components carry
equal incident intensity, giving
\begin{equation}
    T_{\mathrm{lin}}
    =
    \frac{1}{2}
    \left(
        T_{\mathrm{RCP}}
        +
        T_{\mathrm{LCP}}
    \right).
    \label{eq:TMM_linear}
\end{equation}

\begin{figure}[!htbp]
    \centering

    \begin{minipage}[t]{0.485\textwidth}
        \vspace{0pt}
        \makebox[\linewidth][l]{\textbf{\large a}}
        \vspace{-2mm}

        \centering
        \includegraphics[
            width=\linewidth
        ]{chiral_transmission.pdf}
    \end{minipage}%
    \hfill%
    \begin{minipage}[t]{0.485\textwidth}
        \vspace{0pt}
        \makebox[\linewidth][l]{\textbf{\large b}}
        \vspace{-2mm}

        \centering
        \includegraphics[
            width=\linewidth
        ]{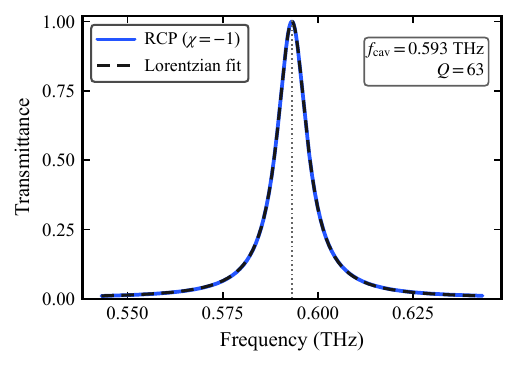}
    \end{minipage}

    \vspace{3mm}

    \begin{minipage}[t]{0.58\textwidth}
        \vspace{0pt}
        \makebox[\linewidth][l]{\textbf{\large c}}
        \vspace{-2mm}

        \centering
        \includegraphics[
            width=\linewidth
        ]{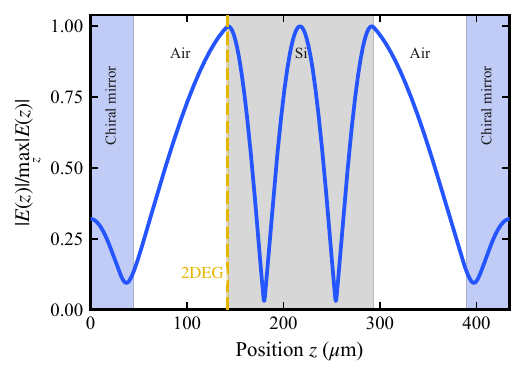}
    \end{minipage}

    \caption{
    \textbf{Passively loaded cavity characterization.}
    \textbf{a}, TMM transmittance for RCP ($\chi=-1$), LCP
    ($\chi=+1$) and linearly polarized incidence, including the passive
    $2.4~\mu\mathrm{m}$ QW membrane. The cavity supports an isolated RCP
    resonance near $0.593~\mathrm{THz}$ while the LCP response is strongly
    suppressed.
    \textbf{b}, Lorentzian fit to the RCP cavity resonance used to extract
    the resonance frequency and quality factor.
    \textbf{c}, RCP electric-field amplitude reconstructed from the TMM at
    the loaded-cavity resonance. The yellow dashed line indicates the
    effective 2DEG position at the center of the QW membrane.
    }
    \label{fig:S_empty_cavity}
\end{figure}

Figure~\ref{fig:S_empty_cavity} characterizes the passively loaded cavity.
The RCP channel supports an isolated cavity resonance near
$0.593~\mathrm{THz}$, whereas the LCP channel is strongly suppressed.
The RCP resonance is fitted by a Lorentzian profile to determine its central
frequency and quality factor. The corresponding electric-field profile is
reconstructed directly from the forward- and backward-propagating amplitudes
obtained from the TMM.

The effective 2DEG position is taken at the center of the QW membrane,
\begin{equation}
    z_{\mathrm{2DEG}}
    =
    d_{\mathrm M}
    +
    d_{\mathrm{air}}
    +
    \frac{d_{\mathrm{QW}}}{2}
    =
    142.2~\mu\mathrm{m},
    \label{eq:TMM_z2deg}
\end{equation}
close to an antinode of the selected RCP cavity field.


\FloatBarrier


\subsection*{Cavity coupled to the 2DEG}

We now activate the free-carrier magneto-optical response of the same
multiple-quantum-well membrane while leaving its passive dielectric background
unchanged. The parameters used in the calculation are:
\begin{equation}
    N_{\mathrm{QW}}=10,
    \qquad
    d_{\mathrm{QW}}=2.4~\mathrm{\mu m},
    \qquad
    n_{\mathrm{2D}}
    =
    3.2\times10^{12}~\mathrm{cm}^{-2},
    \label{eq:TMM_QW1}
\end{equation}
and
\begin{equation}
    m^\ast=0.067m_0,
    \qquad
    \epsilon_{\mathrm{bg}}=12.96,
    \qquad
    \frac{\gamma}{2\pi}=5.7~\mathrm{GHz}.
    \label{eq:TMM_QW2}
\end{equation}
These parameters are representative of GaAs Landau-polariton systems used in
polarization-resolved terahertz spectroscopy~\cite{Li2018}.

The positive cyclotron frequency is
\begin{equation}
    \omega_c
    =
    \frac{e|B|}{m^\ast}.
    \label{eq:TMM_wc}
\end{equation}
In the circular basis, the active and inactive cyclotron responses are
~\cite{Li2018}
\begin{equation}
    \epsilon_{\mathrm{CRA}}(\omega,B)
    =
    \epsilon_{\mathrm{bg}}
    -
    \frac{
        \omega_p^2
    }{
        \omega
        \left(
            \omega-\omega_c+i\gamma
        \right)
    },
    \label{eq:TMM_CRA}
\end{equation}
and
\begin{equation}
    \epsilon_{\mathrm{CRI}}(\omega,B)
    =
    \epsilon_{\mathrm{bg}}
    -
    \frac{
        \omega_p^2
    }{
        \omega
        \left(
            \omega+\omega_c+i\gamma
        \right)
    },
    \label{eq:TMM_CRI}
\end{equation}
where
\begin{equation}
    \omega_p^2
    =
    \frac{
        e^2n_{\mathrm{2D}}
    }{
        \epsilon_0m^\ast d_{\mathrm{QW}}
    }.
    \label{eq:TMM_wp}
\end{equation}

The corresponding refractive index of the QW layer is
\begin{equation}
    n_{\mathrm{QW},\chi}(\omega,B)
    =
    \sqrt{
        \epsilon_\chi(\omega,B)
    },
    \label{eq:TMM_nQW}
\end{equation}
and its characteristic matrix is obtained from
Eq.~\eqref{eq:TMM_layer_matrix} in exactly the same way as for the other
layers. The complete cavity--2DEG transfer matrix is therefore
\begin{equation}
    \mathcal M_\chi^{\mathrm{2DEG}}
    =
    \mathcal M_{M_\chi}
    \mathcal M_{\mathrm{air}}
    \mathcal M_{\mathrm{QW},\chi}
    \mathcal M_{\mathrm{Si}}
    \mathcal M_{\mathrm{air}}
    \mathcal M_{M_\chi}.
    \label{eq:TMM_total_2DEG}
\end{equation}
The transmission is then obtained from the same
Eqs.~\eqref{eq:TMM_t} and \eqref{eq:TMM_T}.

Reversal of the magnetic field exchanges which circular polarization experiences the CRA
or CRI response:
\begin{equation}
\begin{array}{c|cc}
    &
    \mathrm{RCP}\;(\chi=-1)
    &
    \mathrm{LCP}\;(\chi=+1)
    \\[1mm]
    \hline
    B<0
    &
    \mathrm{CRA}
    &
    \mathrm{CRI}
    \\[1mm]
    B>0
    &
    \mathrm{CRI}
    &
    \mathrm{CRA}.
\end{array}
\label{eq:TMM_mapping}
\end{equation}
The chiral-mirror response itself is not changed when the magnetic field is
reversed.

\begin{figure}[!htbp]
    \centering


    \begin{minipage}[t]{0.485\textwidth}
        \vspace{0pt}
        \makebox[\linewidth][l]{\textbf{\large a}}
        \vspace{-2mm}

        \centering
        \includegraphics[
            width=\linewidth
        ]{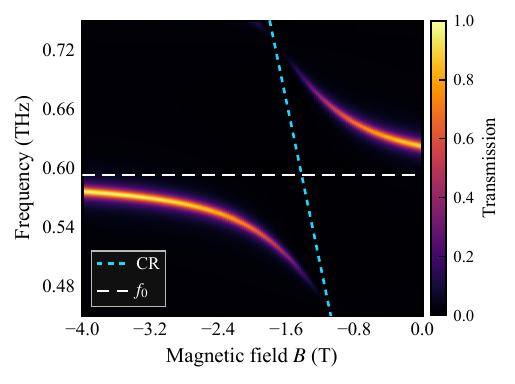}
    \end{minipage}%
    \hfill%
    \begin{minipage}[t]{0.485\textwidth}
        \vspace{0pt}
        \makebox[\linewidth][l]{\textbf{\large b}}
        \vspace{-2mm}

        \centering
        \includegraphics[
            width=\linewidth
        ]{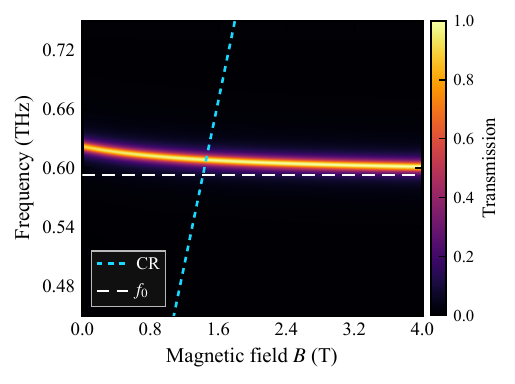}
    \end{minipage}


    \begin{minipage}[t]{0.485\textwidth}
        \vspace{0pt}
        \makebox[\linewidth][l]{\textbf{\large c}}
        \vspace{-2mm}

        \centering
        \includegraphics[
            width=\linewidth
        ]{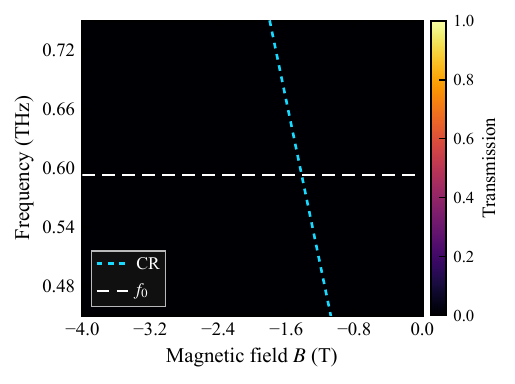}
    \end{minipage}%
    \hfill%
    \begin{minipage}[t]{0.485\textwidth}
        \vspace{0pt}
        \makebox[\linewidth][l]{\textbf{\large d}}
        \vspace{-2mm}

        \centering
        \includegraphics[
            width=\linewidth
        ]{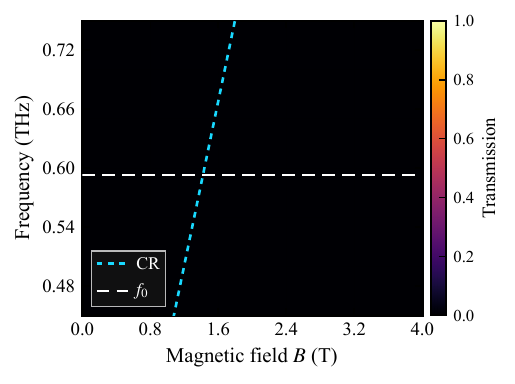}
    \end{minipage}


    \begin{minipage}[t]{0.485\textwidth}
        \vspace{0pt}
        \makebox[\linewidth][l]{\textbf{\large e}}
        \vspace{-2mm}

        \centering
        \includegraphics[
            width=\linewidth
        ]{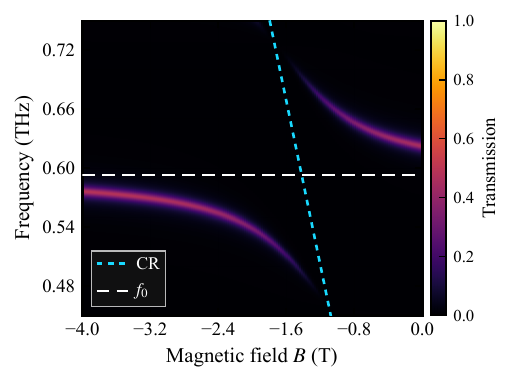}
    \end{minipage}%
    \hfill%
    \begin{minipage}[t]{0.485\textwidth}
        \vspace{0pt}
        \makebox[\linewidth][l]{\textbf{\large f}}
        \vspace{-2mm}

        \centering
        \includegraphics[
            width=\linewidth
        ]{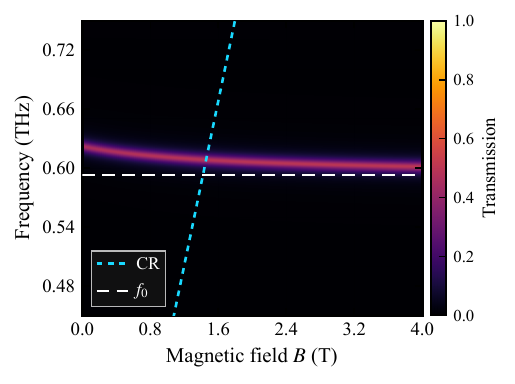}
    \end{minipage}

    \caption{
    \textbf{Polarization-resolved magnetotransmission of the
    cavity--2DEG system.}
    \textbf{a,b}, RCP ($\chi=-1$) transmission for $B<0$ and $B>0$,
    respectively. For $B<0$ the RCP cavity mode probes the CRA response and
    develops an avoided crossing; for $B>0$ it probes the CRI response.
    \textbf{c,d}, LCP ($\chi=+1$) transmission for the corresponding field
    directions. The LCP cavity channel remains strongly suppressed.
    \textbf{e,f}, Transmission for linearly polarized incidence,
    $T_{\mathrm{lin}}=(T_{\mathrm{RCP}}+T_{\mathrm{LCP}})/2$.
    The cyan dashed line denotes the bare cyclotron resonance and the white
    dashed line the bare cavity resonance.
    }
    \label{fig:S_TMM_maps}
\end{figure}

The six polarization-resolved calculations are shown in
Fig.~\ref{fig:S_TMM_maps}. For $B<0$, the RCP cavity field probes the CRA
response and develops the conventional upper- and lower-polariton avoided
crossing. For $B>0$, the same RCP cavity mode probes the CRI response and the
conventional anticrossing disappears. The LCP cavity response remains strongly
suppressed for both field directions. The linear-probe response consequently
inherits the same asymmetry through its RCP component.


\FloatBarrier


\subsection*{Cavity vacuum-field profile and collective coupling}

The cavity mode entering the quantum Hamiltonian is obtained from the
empty-cavity RCP electric field evaluated at the bare cavity resonance
$\omega_0$. We denote the complex electric field reconstructed from the
transfer-matrix calculation by $E_{\mathrm{TMM}}(z)$. Since the transfer
matrix method is linear, the incident field amplitude is arbitrary and is
set to unity; only the relative spatial profile is relevant for the mode
normalization.

The quantized vacuum electric field can be written as
\begin{equation}
    E_{\mathrm{vac}}(z)
    =
    \sqrt{
        \frac{
            \hbar\omega_0
        }{
            2\epsilon_0 S
        }
    }
    \frac{
        E_{\mathrm{TMM}}(z)
    }{
        \left[
            \displaystyle
            \int dz\,
            \epsilon(z)
            |E_{\mathrm{TMM}}(z)|^2
        \right]^{1/2}
    },
    \label{eq:TMM_Evac}
\end{equation}
where $S$ is the transverse effective mode area. The corresponding
one-dimensional vacuum-field amplitude is therefore conveniently expressed
as
\begin{equation}
    \boxed{
    \sqrt{S}\,
    E_{\mathrm{vac}}(z)
    =
    \sqrt{
        \frac{
            \hbar\omega_0
        }{
            2\epsilon_0
        }
    }
    \frac{
        E_{\mathrm{TMM}}(z)
    }{
        \left[
            \displaystyle
            \int dz\,
            \epsilon(z)
            |E_{\mathrm{TMM}}(z)|^2
        \right]^{1/2}
    }.
    }
    \label{eq:TMM_Evac_sqrtS}
\end{equation}
This quantity is independent of both the arbitrary TMM input amplitude and
the unknown transverse mode area $S$.

\begin{figure}[!htbp]
    \centering

    \includegraphics[
        width=0.58\linewidth
    ]{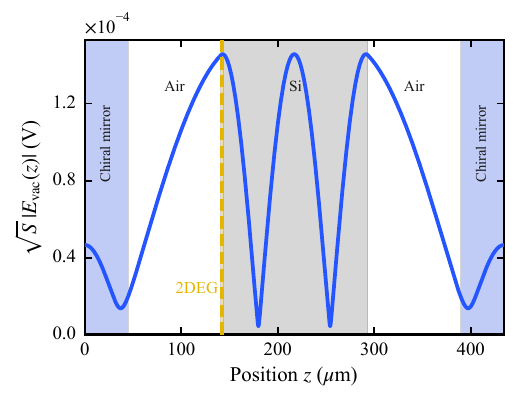}

    \caption{
    \textbf{RCP cavity vacuum electric-field profile.}
    Absolute value of the one-dimensional vacuum electric-field amplitude
    $\sqrt{S}\,|E_{\mathrm{vac}}(z)|$, obtained from the empty-cavity TMM
    field through Eq.~\eqref{eq:TMM_Evac_sqrtS}. The yellow dashed line
    indicates the effective 2DEG position used to evaluate the collective
    cyclotron coupling.
    }
    \label{fig:S_mode_profile}
\end{figure}

Figure~\ref{fig:S_mode_profile} shows the resulting RCP vacuum electric-field
profile. In contrast to the energy-normalized auxiliary mode function
$\sqrt{\epsilon(z)}E(z)$, the physical tangential electric field remains
continuous across dielectric interfaces. At the effective 2DEG position,
the present cavity gives
\begin{equation}
    \boxed{
    \sqrt{S}\,
    \left|
        E_{\mathrm{vac}}(z_{\mathrm{2DEG}})
    \right|
    =
    1.4564\times10^{-4}\ \mathrm{V}.
    }
    \label{eq:Evac_2DEG}
\end{equation}

The collective cyclotron coupling can be written directly in terms of this
vacuum-field amplitude. For a cyclotron frequency $\omega_c(B)$ and a
cavity mode of frequency $\omega_0$,
\begin{equation}
    \bar g(B)
    =
    e
    \sqrt{
        \frac{
            n_{\mathrm{2D}}\omega_c(B)
        }{
            m^\ast\hbar\omega_0^2
        }
    }
    \,
    \sqrt{S}\,
    \left|
        E_{\mathrm{vac}}(z_{\mathrm{2DEG}})
    \right|.
    \label{eq:TMM_gbar_Evac}
\end{equation}
At cyclotron resonance,
\begin{equation}
    \omega_c=\omega_0,
\end{equation}
this reduces to
\begin{equation}
    g
    =
    e
    \sqrt{
        \frac{
            n_{\mathrm{2D}}
        }{
            m^\ast\hbar\omega_0
        }
    }
    \,
    \sqrt{S}\,
    \left|
        E_{\mathrm{vac}}(z_{\mathrm{2DEG}})
    \right|.
    \label{eq:TMM_g}
\end{equation}

Using
$n_{\mathrm{2D}}=3.2\times10^{12}\,\mathrm{cm}^{-2}$ together with the
cavity vacuum-field amplitude of Eq.~\eqref{eq:Evac_2DEG} gives
\begin{equation}
    \boxed{
        \frac{g}{\omega_0}
        =
        0.2287
        \simeq
        0.23 .
    }
    \label{eq:TMM_g_ratio}
\end{equation}

Equivalently, away from resonance the field-dependent coupling can be
expressed in terms of the resonant value $g$ as
\begin{equation}
    \boxed{
    \bar g(B)
    =
    g
    \sqrt{
        \frac{
            \omega_c(B)
        }{
            \omega_0
        }
    }.
    }
    \label{eq:TMM_gbar}
\end{equation}


\FloatBarrier

\subsection*{Supplementary Note 5: Full-wave finite-element validation}

To independently validate the transfer-matrix results, we performed full-wave
finite-element simulations in COMSOL Multiphysics using the same cavity geometry
and material parameters as in Supplementary Note 4.
COMSOL uses the time convention $e^{+i\omega t}$. To preserve the physical
RCP/LCP polarization convention used throughout the manuscript,
$\chi=-1\equiv\mathrm{RCP}$ and $\chi=+1\equiv\mathrm{LCP}$, we define
\[
\mathbf e_\chi=
\frac{\mathbf e_x-i\chi\mathbf e_y}{\sqrt2}.
\]
The magneto-optical response of the QW membrane is implemented through the
circular dielectric eigenvalues
\[
\epsilon_\chi(\omega,B)
=
\epsilon_{\mathrm{bg}}
-
\frac{\omega_p^2}
{\omega\left(\omega-\chi\omega_c-i\gamma\right)},
\]
where
\[
\omega_c=\frac{eB}{m^\ast}.
\]

\begin{figure}[!t]
    \centering
    \includegraphics[width=0.62\linewidth]{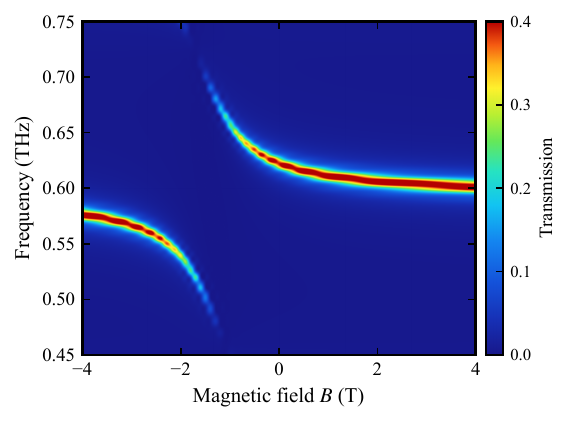}
    \caption{
    \textbf{Full-wave finite-element magnetotransmission.}
    COMSOL transmission spectrum of the chiral cavity--2DEG system as a
    function of magnetic field and frequency. For $B<0$, the selected RCP
    cavity mode undergoes resonant hybridization with the cyclotron-active
    response and forms an avoided crossing. For $B>0$, the same cavity mode
    probes the cyclotron-inactive response and the conventional polaritonic
    anticrossing disappears.
    }
    \label{fig:comsol_map}
\end{figure}

The corresponding Cartesian gyrotropic tensor is obtained by transforming these
two circular eigenvalues to the $(x,y)$ basis. The chiral-mirror response is kept
fixed throughout the magnetic-field sweep, with the RCP channel transmitted and
the LCP channel strongly suppressed, exactly as in the transfer-matrix model.

The resulting full-wave magnetotransmission is shown in Fig.~\ref{fig:comsol_map}.
For $B<0$, the RCP cavity mode couples to the cyclotron-active response and
develops the conventional upper- and lower-polariton avoided crossing.
For $B>0$, the same fixed cavity mode probes the cyclotron-inactive response,
and the conventional anticrossing disappears, leaving the displaced
cavity-like resonance. The full-wave calculation therefore independently
reproduces the asymmetric field-dependent response obtained from the TMM.

\nolinenumbers
\renewcommand{\refname}{Supplementary References}
\putbib[bibliography]
\end{bibunit}

\end{document}